%% file: main.tex
\documentclass[manuscript,screen,authorversion,nonacm]{acmart}

\usepackage[toc,page]{appendix} 
\usepackage{multirow}
 \usepackage{array}

\usepackage{longtable}
\usepackage{hyperref}
\usepackage{pdfpages}
\usepackage{rotating}

\usepackage{multirow}
\usepackage{array}
\usepackage{graphicx}
\usepackage{pdflscape}

\AtBeginDocument{%
  }

\usepackage{comment}
\usepackage{tcolorbox}
\tcbuselibrary{skins}
\newtcolorbox{txtbox}{%
enhanced jigsaw,drop shadow=darkgray!50!white,colback=white,opacityframe=0.5}

\usepackage{draftwatermark}
\SetWatermarkText{Just Accepted}
\SetWatermarkScale{0.5}

\begin{document}

\title{Exploring Empathy in Software Engineering: Insights from a Grey Literature Analysis of Practitioners' Perspectives}
\thanks{\textbf{This is the author’s version of the paper accepted for publication in ACM Transactions on Software Engineering and Methodology (TOSEM). The final version will be available via the ACM Digital Library.}}



\author{Lidiany Cerqueira}
\orcid{0000-0002-4989-0986}
\email{lidiany.cerqueira@ufba.br}
\affiliation{%
  \institution{Federal University of Bahia}
  \city{Salvador}
  \state{Bahia}
    \country{Brazil}    
      }
      \affiliation{%
  \institution{Virginia Commonwealth University}
  \city{Richmond}
  \state{Virginia}
  \country{USA}
}

\author{João Pedro Bastos}
\orcid{0009-0004-4247-1430}
\email{jpsbastos@ecomp.uefs.br}
\affiliation{%
 \institution{State University of Feira de Santana}
     \city{Feira de Santana}
     \state{Bahia}
      \country{Brazil}
}
\affiliation{%
  \institution{University of Porto}
  \city{Porto}
  \country{Portugal}
}

 \author{Danilo Neves}
\orcid{0000-0003-4754-7882}
\email{danilo.neves@ifs.edu.br}
\affiliation{%
  \institution{Federal Institute of Sergipe}
  \city{Lagarto}
   \state{Sergipe}
    \country{Brazil}
      } 

\author{Glauco Carneiro}
\orcid{0000-0001-6241-1612}
\email{glauco.carneiro@dcomp.ufs.br}
\affiliation{%
  \institution{Federal University of Sergipe}
  \city{São Cristóvão}
     \state{Sergipe}
      \country{Brazil}
      }
 
\author{Rodrigo Spínola}
\orcid{0000-0003-0272-9578}
\email{spinolaro@vcu.edu}
\affiliation{%
     \institution{Virginia Commonwealth University}     
      \city{Richmond}
    \state{Virginia}
      \country{United States of America}
  }

 \author{Sávio Freire}
\orcid{0000-0002-3989-9442}
\email{savio.freire@ifce.edu.br}
\affiliation{%
 \institution{Federal Institute of Ceará}
 \city{Morada Nova}
     \state{Ceará}
      \country{Brazil}
 } 

\author{José Amancio Macedo Santos}
\orcid{0000-0002-9509-5238}
\email{zeamancio@uefs.br}
\affiliation{%
\institution{State University of Feira de Santana}
   \city{Feira de Santana}
     \state{Bahia}
      \country{Brazil}
    }
    
\author{Manoel Mendonça}
\orcid{0000-0002-0874-7665}
\email{manoel.mendonca@ufba.br}
\affiliation{%
  \institution{Federal University of Bahia}
  \city{Salvador}
     \state{Bahia}
      \country{Brazil}
      }

\renewcommand{\shortauthors}{Cerqueira et al.}
\acmArticleType{Review}



\begin{abstract}
\textbf{Context}. Empathy, a key social skill, is essential for communication and collaboration in SE but remains an under-researched topic. \textbf{Aims}. This study investigates empathy in SE from practitioners’ perspectives, aiming to characterize its meaning, identify barriers, discuss practices to overcome them, and explore its effects. \textbf{Method}. A qualitative content analysis was conducted on 55 web articles from DEV and Medium, two communities widely used by practitioners. To strengthen our findings, we conducted a follow-up survey with empathy experts. \textbf{Results}. The study proposes a definition of empathy in SE, identifies barriers such as toxic culture and excessive technical focus, practices to foster empathy in teams, and outcomes, including improved collaboration, communication, and reduced anxiety, frustration, and stress. These findings are synthesized into a conceptual framework. \textbf{Conclusion}. Survey results indicate the framework is clear, valuable, and raises empathy awareness, with suggestions for improvements and integration into training. This study paves the way for improving team dynamics by addressing barriers and offering strategies to cultivate empathy. Future work will explore empathy’s broader implications in SE practice.
\end{abstract}



\maketitle

\section{Introduction}
\label{sec:introduction}

\input{content/1_introduction} 

\section{Background}
\label{sec:background}

\input{content/2_background}

\section{Research strategy} 
\label{sec:research}
\input{content/3_research_method}

\section{Results}
\label{sec:results}
\input{content/4_results}

\section{Discussion}
\label{sec:discussion}
\input{content/5_discussion}

\input{content/5.1_framework}

\input{content/5.2_evaluating_the_framework}

\section{Limitations}
\label{sec:limitations}

\input{content/5.4_limitations}

\section{Conclusion}
\label{sec:conclusion}
\input{content/6_conclusion}

\begin{acks} 
We thank the editors and anonymous reviewers for their thoughtful feedback, which helped improve this work. We are grateful to the practitioners who shared their insights through Medium and DEV communities and answered the survey for their valuable contributions. We also thank the respondents of the pilot survey for their valuable time and feedback.

This study was financed in part by the Coordenação de Aperfeiçoamento de Pessoal de Nível Superior Brasil (CAPES), Finance Code 001, and by the Conselho Nacional de Desenvolvimento Científico e Tecnológico (CNPq).
\end{acks}

\bibliographystyle{unsrtnat}
\bibliography{refs}
\appendix
\input{content/7_appendix_search}
\input{content/7.1_appendix_reliability}

\input{content/7.2_appendix_extraction}
\input{content/7.3_appendix_results}

\input{content/7.4_appendix_survey}

\end{document}

%% file: content/1_introduction.tex
Empathy is among the most in-demand skills at work\footnote{Data from the World Economic Forum: \url{https://www.weforum.org/publications/the-future-of-jobs-report-2025/in-full/3-skills-outlook/}}. It is the ability to understand and vicariously experience the emotions of others~\cite{decety2021emergence}. It allows understanding others' emotions, feelings, and perspectives \cite{gladstein1983understanding}. Empathy promotes social behaviors \cite{lockwood2016anatomy} and is one of the skills necessary for effective communication \cite{riggio2008emotional}. 

In the SE context, empathy is one of the aspects influencing software development activities~\cite{cerqueira2023sbes}, it is critical for many software development roles \cite{acuna2006emphasizing}, a required non-technical skill in the software development market \cite{rabelo2022role}, and a key factor in improving workplace culture and retention, particularly for women in software teams \cite{trinkenreich2022empirical}. However, despite its value for software practitioners, it remains an under-researched topic in SE \cite{gunatilake2023}. Regardless of growing interest in the impact of empathy in SE, few studies have explored it \cite{cerqueira2023sbes,gunatilake2024enablers}. Due to its multifaceted nature, one of the challenges is outlining empathy in this context \cite{devathasan2024deciphering}. 

Software practitioners widely discuss SE topics, such as non-technical skills, in Grey Literature (GL)~\cite{papoutsoglou2021mining}. GL is valuable in SE as it may contain practical insights, case studies, and the latest developments that may not yet be captured in peer-reviewed journals~\cite{garousi2019guidelines}. They supply a rich source of information for researchers, bringing to light the practitioners' point of view \cite{gomes2023investigating}. The sources of GL are an informal knowledge base that complements formal research, helping practitioners and researchers stay informed about the latest trends, tools, and best practices in the field. The GL, based on online communities, well represents the voice of practitioners in terms of public discussion \cite{kamei2021evidence}. 

We have investigated empathy in our previous studies~\cite{cerqueira2023sbes,cerqueira2024empathy} by considering 22 articles from the DEV community\footnote{DEV (\url{https://dev.to/}) is an online community for software developers in which they write long-form posts about their experiences, preferences, and working lives.}, revealing the meaning of empathy for software practitioners,  empathetic practices, and their effects on software projects.
In this current study, we significantly expanded our dataset by including 33 articles from Medium\footnote{Medium is a prominent source of GL where practitioners share experiences and discuss technical challenges~\url{https://medium.com/}.}.  
By synthesizing the analysis of these 55 web articles, this study provides a broader investigation into practitioners’ perceptions of empathy in SE. Moreover, we extended our analysis to examine barriers software practitioners face to sustain empathy in workplace settings across the entire dataset. Thus, we performed a detailed content analysis \cite{krippendorff2018content} of 55 web articles to answer the following research questions (RQs): 

\begin{itemize} 
    \item RQ1: What does empathy mean for software practitioners?
    \item RQ2: What are the barriers to empathy in software engineering?
    \item RQ3: How to practice empathy in software engineering?
    \item RQ4: What are the effects of practicing empathy in software engineering?
\end{itemize}   

By addressing these questions, we aim to bridge the gap between research and practice, define empathy in SE, understand the barriers that hinder empathy in SE workplaces, and build a set of empathetic practices in SE and their expected effects. In addition, we surveyed empathy experts\footnote{The practitioners self-identified as having intermediate to expert-level understanding of empathy in SE.} to assess our findings.

The main contributions of this paper are:

\begin{itemize}
    \item A starting point to define empathy in SE: our work provides a foundational definition of empathy suitable to the SE context. Given that various authors ambiguously define empathy and it remains under-explored in SE, these results help establish a more precise understanding and foster further research on this critical human factor.
    \item A list of barriers to empathy in the workplace: we identify the specific barriers software practitioners face in sustaining empathy within the SE context. 
    \item A discussion of strategies to overcome these barriers: understanding them is crucial for enabling practitioners to develop effective strategies to overcome them, fostering a work environment open to building an empathic culture.
    \item A set of empathetic practices for software practitioners: we provide a set of actionable practices aimed at helping software practitioners develop their empathic capacities.
    \item A set of anticipated outcomes of these practices, including improved team collaboration, enhanced communication, psychological safety, and less blame and stress among practitioners. These anticipated benefits highlight the developmental potential of empathy in fostering healthier and more effective SE work environments.
    \item A conceptual framework of empathy in SE: we present a framework that connects the meaning of empathy, the proposed practices, and their anticipated effects. 
    \item Discussing how practitioners can explore the conceptual framework. It is a valuable resource for researchers and practitioners, offering a structured approach to integrating empathy into SE processes and fostering its application in real-world contexts.    
\end{itemize}

The remainder of this paper is organized as follows. Section \ref{sec:background} presents a literature review on the topic. Section \ref{sec:research} details the research method.
Section \ref{sec:results} explains our findings. Section \ref{sec:discussion} discusses the results. Section \ref{sec:limitations} presents the limitations of our research. Section \ref{sec:conclusion} concludes the paper and proposes future work.

%% file: content/2_background.tex
In this study, we draw on interdisciplinary research to build a background for our work. As SE researchers advocate for employing methodologies from other behavioral sciences to explore the human dimensions of SE \cite{lenberg2015behavioral}, we start by gathering how neuroscience and psychology researchers define empathy. Next, we present related work that addresses empathy in SE. 

 Empathy is multifaceted and complex \cite{davis2018empathy}. It is often confused with related concepts such as sympathy. However, psychological researchers argue that these terms are not synonymous \cite{wispe1986distinction}. Each term stems from different historical and theoretical backgrounds and reflects distinct emotional and cognitive processes. Clarifying this distinction is essential to understanding how software practitioners engage with empathy in their professional contexts.

\citet{decety2021emergence} proposed a conceptual empathy model based on neuroscience and social sciences findings. This model describes four facets of empathy: emotional empathy, compassion, cognitive empathy, and emotional regulation. \textbf{Emotional empathy} is the ability to share the emotional state of others. \textbf{Compassion} is caring for the well-being of others. \textbf{Cognitive empathy} is the capacity to adopt another person's perspective to understand his/her thinking, feelings, and emotions. \textbf{Emotional regulation} evaluates and modifies one's emotional response to achieve a goal. We adopt the model they proposed as the foundation of our work due to its interdisciplinary strength, bridging neuroscience and social sciences to offer a comprehensive understanding of empathy. 

\subsection{Empathy in Software Engineering}
\label{sec:related_work}
   \citet{acuna2006emphasizing} identified empathy as a necessary interpersonal skill for multiple software roles. \citet{dutra2021human} highlighted empathy as the most significant human factor impacting productivity and the code review process. Similarly, \citet{rabelo2022role} noted empathy as a required non-technical skill in the software development market and emphasized the need for a taxonomy. 

Empathy is particularly critical in Requirements Engineering (RE), helping developers understand users' needs~\cite{ferreira2015eliciting,levy2018importance}. Techniques such as the Empathy Map are used in user experience (UX) and design thinking (DT) to support understanding of user needs and to communicate findings to others on the team, fostering a shared understanding. ~\cite{gasparini2015perspective}. The DT can help achieve empathy by acquiring a cognitive empathic understanding and insight.

\citet{trinkenreich2022empirical} highlighted the importance of empathy and training as a strategy to improve team culture, collaboration, and the retention of women in SE. Their findings, grounded in the lived experiences of practitioners, brought attention to empathy as a key factor in inclusive and effective work environments.

Although these studies did not formalize a model of empathy in SE, they brought visibility to empathy as a socio-organizational concern and paved the way for emerging research that seeks to define, measure, and integrate empathy into SE practice.

\citet{gunatilake2023} proposed a preliminary taxonomy of empathy grounded in multidisciplinary models to increase developers’ understanding of diverse end-user needs. They later identified enablers and barriers of empathy in the developer–user interactions~\cite{gunatilake2024enablers}. However, these studies do not focus on the broader impact of empathy on software projects, teams, and outcomes.

 \citet{levy2024learning} explored how hybrid DT programs foster empathy development among students by examining their skills, orientation, and emotional engagement. They emphasize the importance of empathy in practicing DT and developed a questionnaire to examine empathy skills following \citet{walther2020empathy} empathy model.

\citet{devathasan2024deciphering} examined developer responses to users to identify empathetic engagement. They applied the Perception-Action Model (PAM)~\cite{preston2007perception}, which includes emotion contagion, sympathy, helping behavior, cognitive empathy, and empathetic accuracy. These categories closely align with \citet{decety2021emergence}’s model, which frames empathy as comprising emotional empathy, compassion, cognitive empathy, and emotional regulation. Yet, Devathasan et al. acknowledged the lack of consensus to define empathy in SE.

\subsection{Our contribution}

Despite growing interest in the impact of empathy in SE, many areas remain underexplored ~\cite{gunatilake2023}: the meaning of empathy for practitioners, workplace barriers to empathy, how to cultivate empathic skills, and the effects of empathy in SE. We addressed these gaps in prior studies~\cite{cerqueira2023sbes,cerqueira2024empathy}, analyzing 22 articles from DEV, a Grey Literature (GL)\footnote{Grey Literature (GL) encompasses materials outside traditional academic publishing, such as blog posts, technical documentation, and forum discussions~\cite{garousi2019guidelines}.} source used by developers. In \citet{cerqueira2023sbes}, we explored empathy's meanings, practices, and perceived effects in SE, organizing findings into a preliminary framework. In \citet{cerqueira2024empathy}, we examined how empathy affects developers' well-being and proposed practices to promote mental health at work.  Those studies were the first to address the effects of empathy in SE.
 In this current study, we expand on those results by:
\begin{itemize}
    \item Including 33 additional articles from Medium (another GL source),
    \item Performing a full data analysis,
    \item Broadening the scope to investigate barriers to empathy in the software workplace,
     \item Surveying empathy experts to validate and refine the conceptual framework, besides collecting further insights.
\end{itemize}

\noindent Our contributions include:
\begin{enumerate}
    \item Insights into the meaning of empathy for software practitioners.
    \item An updated list of empathetic practices.
    \item Expected outcomes of these practices.
    \item A comprehensive list of workplace empathy barriers.
    \item Strategies to overcome these barriers.
    \item A conceptual framework encompassing our analysis's full body of knowledge.
     \item A public open experimental package to support replication, reuse, and further research on empathy in SE.
\end{enumerate}

This expanded analysis can guide future research and provide practical support for integrating empathy into SE practice.

%% file: content/3_research_method.tex
\begin{figure}[hb]
\centerline{\includegraphics[scale=0.51, trim=0cm 0.1cm 0cm 0cm, clip]{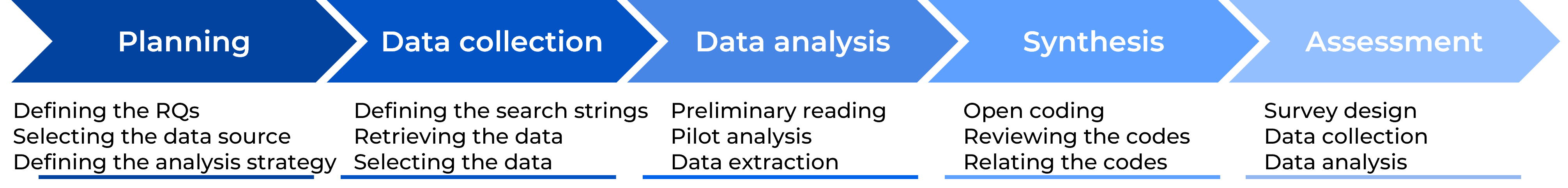}}
\caption{Research steps.}
\label{fig:research_process}
\Description[Overview of the Research Process.]{Overview of the Research Process.}
\end{figure}

Figure~\ref{fig:research_process} presents an overview of the research process. 
It has five main phases: (1) planning, (2) data collection\footnote{To improve readability, we present extended methodological details of the data collection, selection, and extraction in Appendix \ref{appendix_data_collection}.}, (3) data analysis, (4) synthesis, and (5) assessment. In the following sections, we describe these
phases. 

\vspace{-0.5cm}
\subsection{Planning}

In this phase, we establish the research questions, define our research strategy, and the analysis approach adopted in the study.

\subsubsection{Defining the Research Questions}

This study investigates the role of empathy in the context of software development. It focuses on four key dimensions: how empathy is conceptualized, the barriers that inhibit its expression, the practices that foster empathetic behavior, and its effects on software engineering processes and outcomes. To support this investigation, the following research questions (RQs) have been formulated: \label{rqs}

\begin{itemize}
    \item \textbf{RQ1:} \textit{What does empathy mean for software practitioners?} Considering the lack of conceptual clarity regarding the comprehension of empathy from the perspective of software practitioners, this question aims to understand how they define the concept of empathy. ~\citet{decety2020empathy} and ~\citet{hess2016voices} emphasized the importance of investigating conceptual understanding to reduce confusion and ambiguity in the definition of empathy.  Our study aims to understand how software practitioners perceive empathy in their professional practice, particularly in teamwork and collaborative contexts.
   \item \textbf{RQ2:} \textit{What are the barriers to empathy in software engineering?} In workplace environments, there are significant challenges to maintaining empathy across professional and personal spheres. These barriers have been documented in the healthcare field and negatively affect empathy~\cite{banja2006empathy,howick2017overthrowing, jeffrey2019barriers}. This question seeks to identify what barriers software practitioners face to sustain empathy in the SE context.
    \item \textbf{RQ3:} \textit{How to practice empathy in software engineering?} A focus on practices is essential for workers to develop their empathic capacities~\cite{gerdes2009social,walther2017model}. This question aims to identify how software practitioners are practicing empathy. 
   
    \item \textbf{RQ4:} \textit{What are the effects of practicing empathy in software engineering?} This question aims to identify the expected effects of empathetic practices in SE. By answering this question, we can better understand the impact of empathy in SE.
\end{itemize}

\subsubsection{Selecting the data source}

Due to the scarcity of academic research and the lack of a clear taxonomy on empathy in SE \cite{gunatilake2023}, we adopted GL to gather evidence. GL, particularly from online communities, offers access to practitioners' perspectives and has been widely used to explore human aspects in SE \cite{garousi2019guidelines, kamei2021evidence, papoutsoglou2021mining}. Our study follows the research method proposed by \citet{gomes2023investigating}, which focuses on extracting and analyzing practitioner knowledge from online communities.

Following, we explored specialized websites to find sources with SE content, as recommended by \citet{kamei2021grey}. We selected two popular websites among SE practitioners: DEV Community and Medium. 

\textbf{DEV} is an online community for software developers in which they write long-form posts about their experiences, preferences, and working lives. DEV emerged as an essential information hub for software developers and a valuable source to study social aspects of SE \cite{papoutsoglou2021mining}. \textbf{Medium} is a blog platform where professionals write short articles reporting experiences and technical issues. 

Despite GL becoming a popular data source among SE researchers \cite{kamei2021evidence}, previous studies have mainly focused on open source and Q\&A platforms, such as GitHub and Stack Overflow \cite{papoutsoglou2021mining}. However, compared to other online platforms, communities such as DEV and Medium have a more organized and complete structure to introduce and discuss a specific topic \cite{liang2024controlled}. Thus, they are more adequate for our research.

\subsubsection{Defining the analysis strategy}

We adopted content analysis to extract and synthesize data from GL sources
\cite{krippendorff2018content}. Qualitative methods are highly suitable in research addressing human aspects \cite{defranco2017content, lenberg2024qualitative}. Using qualitative methods in SE research can help researchers produce richer and informative results \cite{seaman99}. A qualitative study allows a holistic understanding of empathy by providing an in-depth analysis and gathering diverse participant perspectives.

\subsection{Data analysis}
\label{sec:data_analysis}

\begin{figure}[htb]
\centerline{\includegraphics[scale=0.45, trim=0cm 5cm 0cm 5cm, clip]{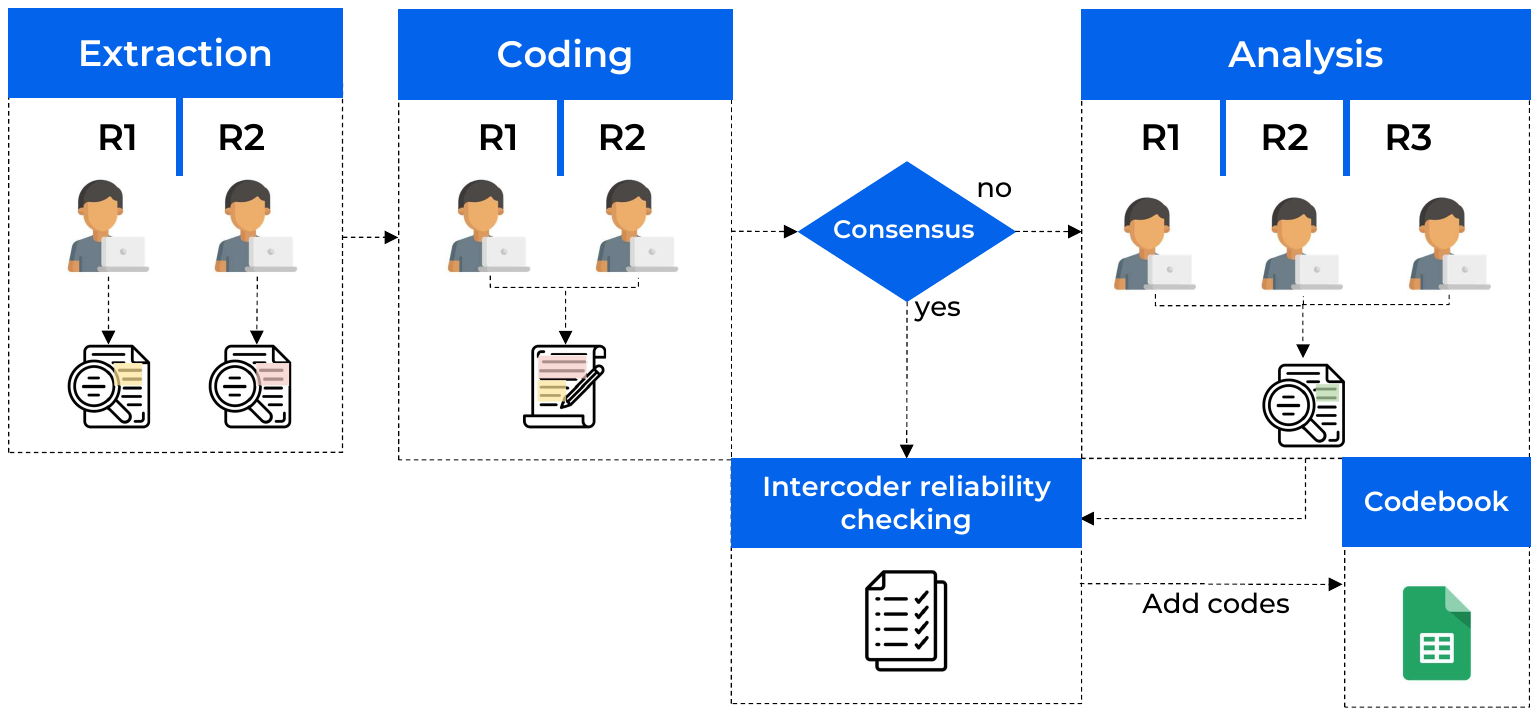}}
\caption{Simplified flowchart of the coding process. Based on the process proposed by \citet{gomes2023investigating}.}
\label{fig:extraction}
\Description[Simplified flowchart of the coding process. Based on the process proposed by \citet{gomes2023investigating}.]{Simplified flowchart of the coding process. Based on the process proposed by \citet{gomes2023investigating}.}
\end{figure}

To address the research questions, we conducted a content analysis following the methodology proposed by \citet{krippendorff2018content}. The overall extraction process is illustrated in Figure \ref{fig:extraction}. Three researchers were involved in this process. Two coders with complementary expertise conducted the analysis\footnote{The first coder is a PhD student with extensive experience in qualitative analysis and experimental SE.}$^,$\footnote{The second coder is an undergraduate research assistant who was specifically hired for this project and received training to perform qualitative analysis.}. A third researcher was involved in the consensus\footnote{The third researcher has a PhD in computer science and extensive experience in empirical SE and qualitative analysis.}. The following sections detail the steps involved in this phase of the study.

\subsubsection{Preliminary reading} 
We conducted a first reading of the articles to familiarize ourselves with the text, get immersed in the data, and seek information to answer the RQs described in Section \ref{rqs}.

\subsubsection{Pilot analysis} \label{sec:pilot}

Due to the complexity of the analysis, we adopted an alignment strategy to ensure the reliability of the extraction. Initially, the two coders conducted a training and alignment step to calibrate their analyses. They started by reading the first three articles independently and comparing the results. With this first step, they sought to guarantee a common understanding of the RQs and the coding criteria. Then, they read the first ten articles and individually answered the RQs. The level of agreement and divergence between the coders was checked in a consensus meeting before analyzing the remaining articles. The third researcher was consulted in case of disagreements. The level of agreement was verified by computing inter-coder reliability measures \cite{lacy2015issues}.

Inter-coder reliability measures are used to assess the degree of agreement between the two coders, evaluate the consistency of the content coding process, guarantee that the coders were on the same page, and ensure the extraction's reliability.

We conducted two rounds of reliability testing: the first during the pilot phase and the second upon completing the whole coding process. Assessing agreement through quantitative measures in collaborative coding is a critical step in qualitative research, as it enhances the rigor and credibility of the findings \cite{gonzalez2023reliability}.

 A detailed discussion of the pilot inter-coder reliability analysis is provided in Appendix~\ref{appendix-reliability}. 
The results of the pilot inter-coder reliability analysis indicate a strong level of agreement between the coders, suggesting that they consistently interpreted and applied the coding scheme similarly across the analyzed articles.

\vspace{-0.1cm}
We detailed the \textbf{data extraction process and synthesis} of our findings in Appendix \ref{appendix-synthesis}, in favor of readability.

\subsection{Assessing the conceptual framework}

From the point of view of empathy experts, we investigated the usefulness, completeness, and accuracy of the proposed conceptual model of empathy in SE. We developed a complementary survey with open-ended and closed questions, ensuring clarity and accessibility. We followed the empirical guidelines to build the questionnaire \cite{ralph2020empirical}. Four senior researchers reviewed the survey in an iterative process. 

\subsubsection{Survey design}
The survey\footnote{The complete survey is available at \url{https://shorturl.at/mbMAH}. It is also on our replication package.} is composed of 28 
questions and is structured in seven sections to facilitate understanding and ensure a logical progression through the topics. Appendix \ref{appendix-survey} details the full questionnaire. 
 
Between January 14, 2025, and January 21, 2025, we piloted the survey with five SE professionals and researchers to review our instrument and pre-test the survey to refine its design. We invited a senior researcher, a doctoral student, and three software practitioners (master's students) to the pilot. In this step, we gauged the time necessary to answer the questionnaire and evaluate if the questions and texts fit the survey goals. The five participants answered the survey in approximately 30 minutes and provided feedback. Based on their feedback, we prepared the final version of the study and distributed it to the participants. 

\subsubsection{Data collection}
To answer the survey, we invited the 51 authors of the articles we collected in DEV and Medium to build the framework, as they are presumed to have specialized knowledge and interest in the topic. We asked them to participate through LinkedIn, directly targeting them based on their contributions to the discourse in the field. The survey was available between January 24, 2025, and March 8, 2025. Participation was voluntary, anonymous, and without incentives.

\subsubsection{Data analysis}
For closed questions and demographic data, we used descriptive statistics. We openly coded the answers for open-ended questions to identify their central ideas~\cite{seaman99}. For example, consider the following answer to Q5: ``\textit{I like specific practices because they are actionable things to actually try, bite-sized}.'' We defined the code \textit{Actionable practices} for it. Two authors performed the coding process. One coded all answers, and the other checked the extracted codes. Divergences were resolved in a consensus meeting.

%% file: content/4_results.tex
This section presents the study's results, beginning with a brief overview of the authors' demographic data to provide context for the findings. Following this, the answer to each research question is addressed in detail.

\subsection{Demographic data}

\begin{figure}[ht]
\centerline{\includegraphics[scale=0.45,trim=0cm 0.5cm 0cm 0.5cm, clip]{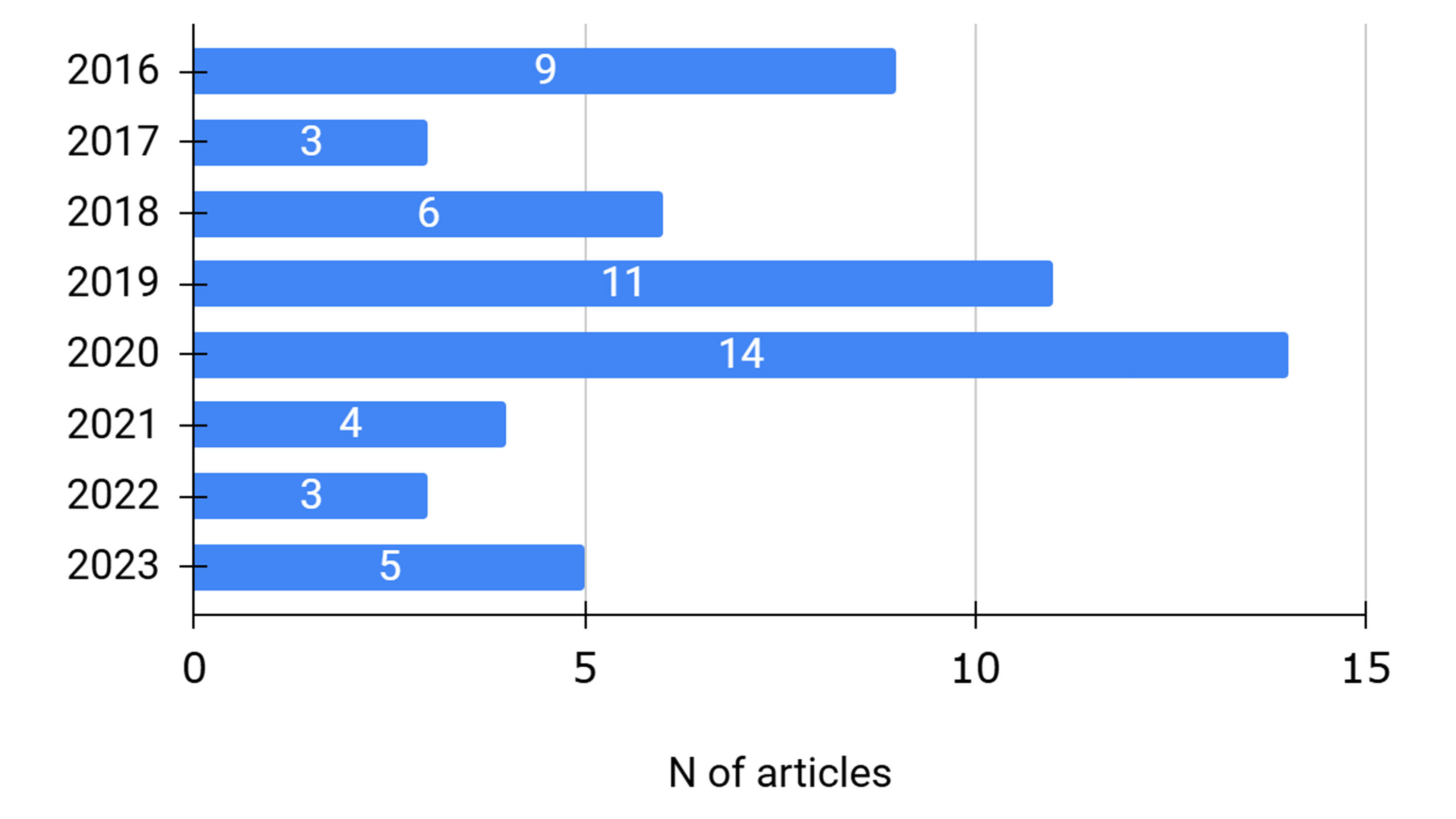}}
\caption{Number of articles selected per year.}
\label{fig:years_chart}
\Description[Number of articles selected per year.]{}
\end{figure}

We analyzed 55 web articles. They had an average of 1,241 words. Although we did not limit the search by a period, the articles were published between 2016 and 2023. 
Figure \ref{fig:years_chart}  shows the distribution per year. We can notice that between 2019 and 2020, the number of published articles about the topic increased compared to the other years. 

\begin{figure}[htb]
\centerline{\includegraphics[scale=0.6, trim=0cm 2cm 0cm 3cm, clip]{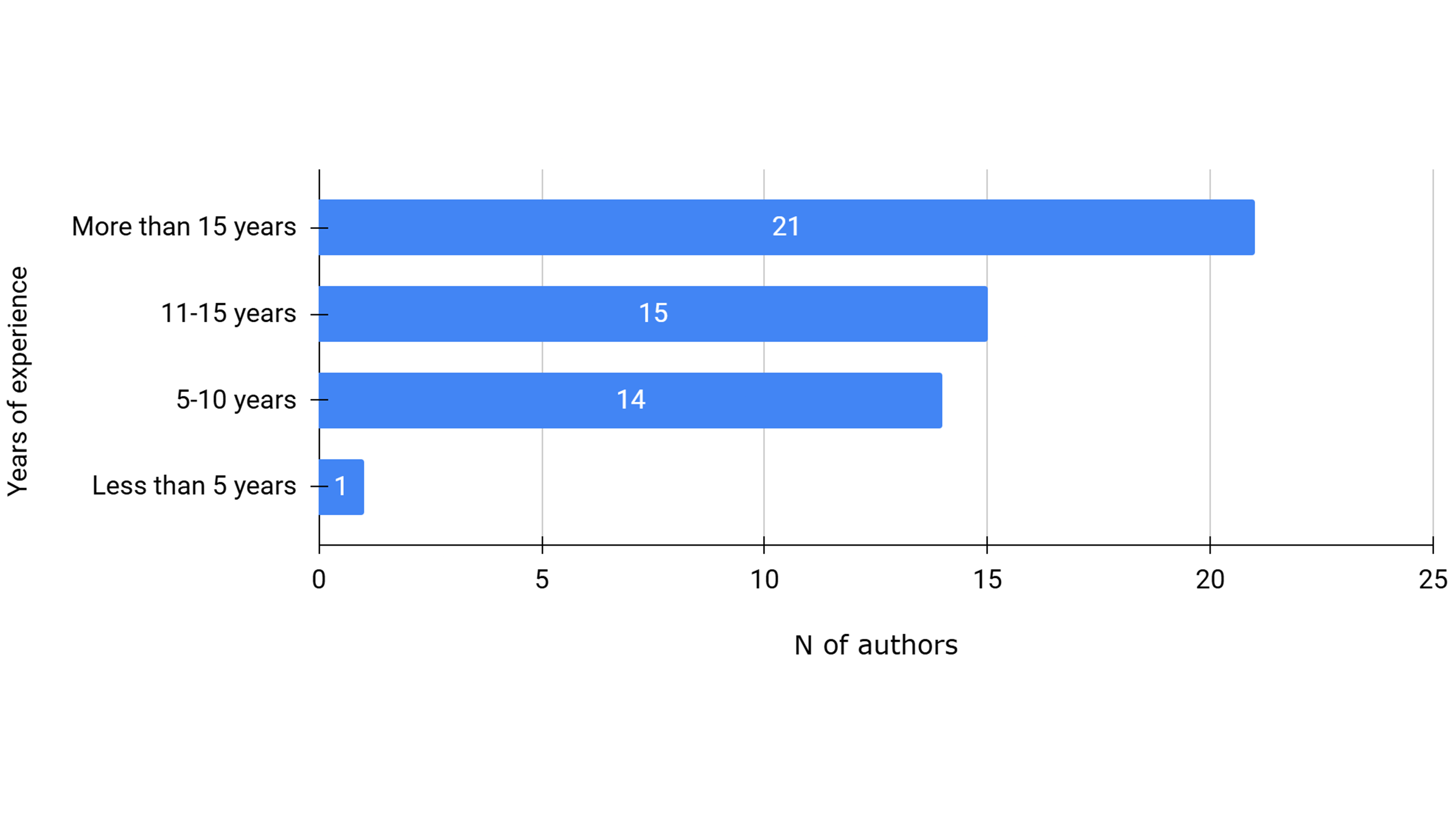}}
\caption{Experience of the authors.}
\label{fig:experience_chart}
\Description[Experience of the authors.]{}
\end{figure}

 We also collected information about the authors of the articles. We retrieved the data from their profiles available on the Medium and Dev platforms, but we also searched their LinkedIn or GitHub profiles to collect demographic data. Some authors wrote more than one article, so the number of authors is 51. To determine their years of experience, we reviewed their profiles, identified the year of their first job, and calculated the duration from that year to the present (2024), providing an estimate of their professional experience. Figure \ref{fig:experience_chart} shows that most authors have more than fifteen years of experience in SE (n = 21).

Figure \ref{fig:experience_chart} shows the primary occupation of the authors. They are in operational roles, such as software engineers and developers (n = 33) and reliability engineers (n = 2), as well as managerial (n = 7), executive (n=2), and developer advocates (n=7) which can be seen as both managerial and operational, depending on the organization and specific job responsibilities. 

 \begin{figure}[htb]
\centerline{\includegraphics[scale=0.5, trim=0cm 3cm 0cm 3cm, clip]{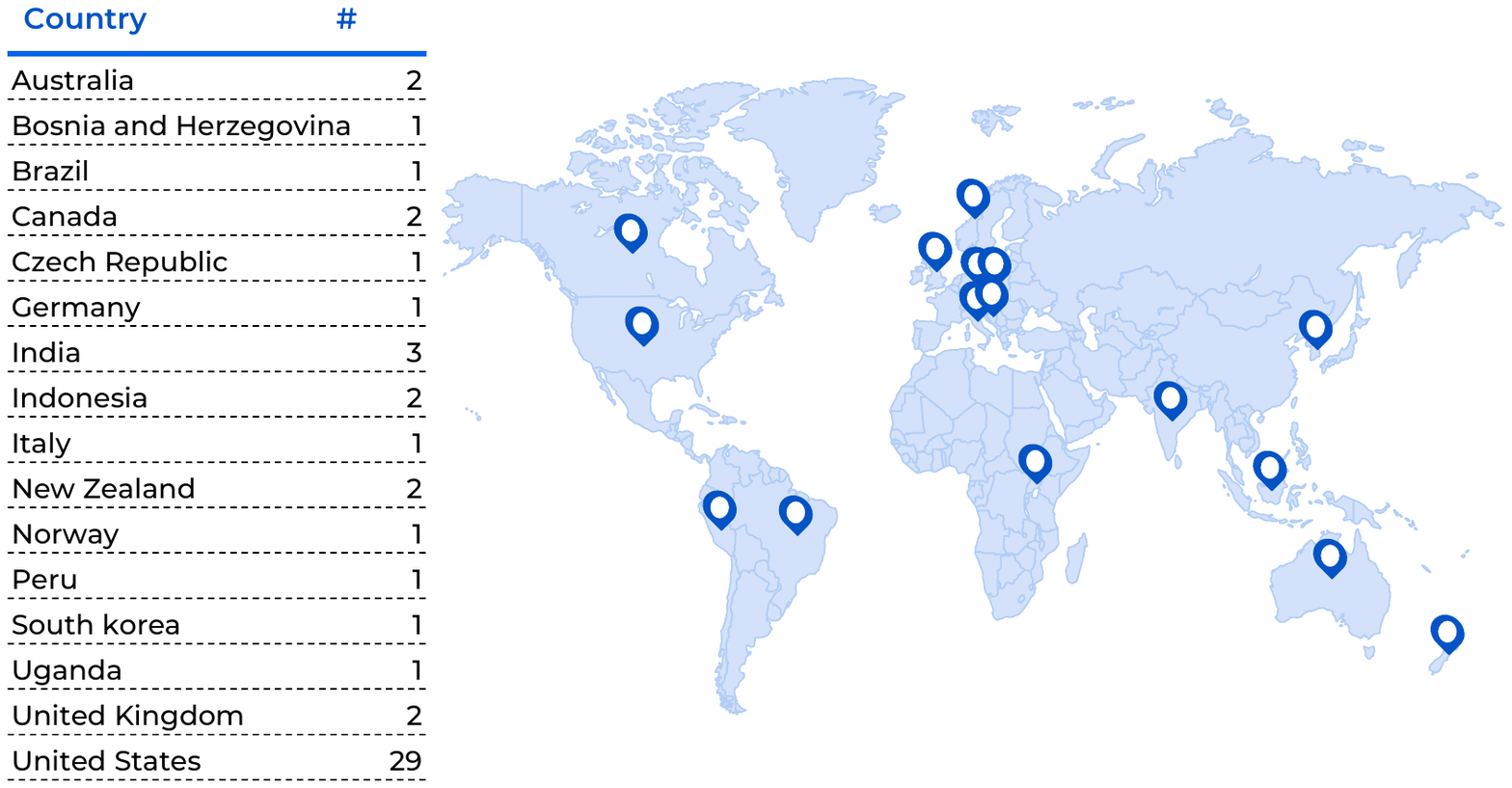}}
\caption{Country distribution of the authors.}
\label{fig:country_chart}
\Description[Country distribution of the authors.]{}
\end{figure}

 Figure \ref{fig:country_chart} shows the frequency of authors per country. We can notice that most authors were from the United States (N = 29). However, there are representations of 15 other countries from different regions of the world, including South America, Asia, Africa, Oceania, and Europe. 

\subsection{RQ1 - What does empathy mean for software practitioners?}

To answer RQ1, we explored practitioners' points of view concerning the meaning of empathy.
Five themes emerged from the analysis of the articles regarding the RQ1: \textit{understanding}, \textit{perspective taking}, \textit{embodiment}, \textit{compassion}, and \textit{emotional sharing}. After coding the themes, we grouped them according to the facets of empathy found in the literature review: cognitive, compassionate, and emotional empathy \cite{decety2020empathy}. In Appendix \ref{sec:app-codes}, Table \ref{tab:facets} presents the codes grouped by facets of empathy, their description, frequency, and the practitioners who mentioned them. 
In the following paragraphs, we discuss them. 

Considering cognitive empathy, there are three themes: understanding, perspective-taking, and embodiment. Most practitioners (n=29) cited \textbf{understanding} as the meaning of empathy for them. It refers to understanding how someone else is thinking or feeling. Practitioners mentioned the capacity to understand another person's thoughts, needs, feelings, problems, and experiences as the meaning of empathy, as we can see in the following quote (P34): ``\textit{[empathy is] the ability to understand how a person feels and what they might be thinking}.''

\textbf{Perspective taking} is the second most cited theme by practitioners (n=18). It means adopting someone else’s subjective perspective. They considered empathy as thinking, considering, imagining, or seeing someone else's perspective. For instance, P39 mentioned it: ``\textit{When discussing empathy...It’s very simply the ability to see things as if from another’s perspective.}''

Following, they cited \textbf{embodiment} as another meaning of empathy (n=11). They described empathy as getting in someone else's shoes or putting oneself in another's place. They stated the need to get in the place of developers, maintainers, clients, users, team members, and commenters. As P40 cited: ``\textit{Empathy is the ability to put yourself in the other person’s shoes.}''

\begin{figure}[htb]
\centerline{\includegraphics[scale=0.6, trim=0cm 2cm 0cm 2cm, clip]{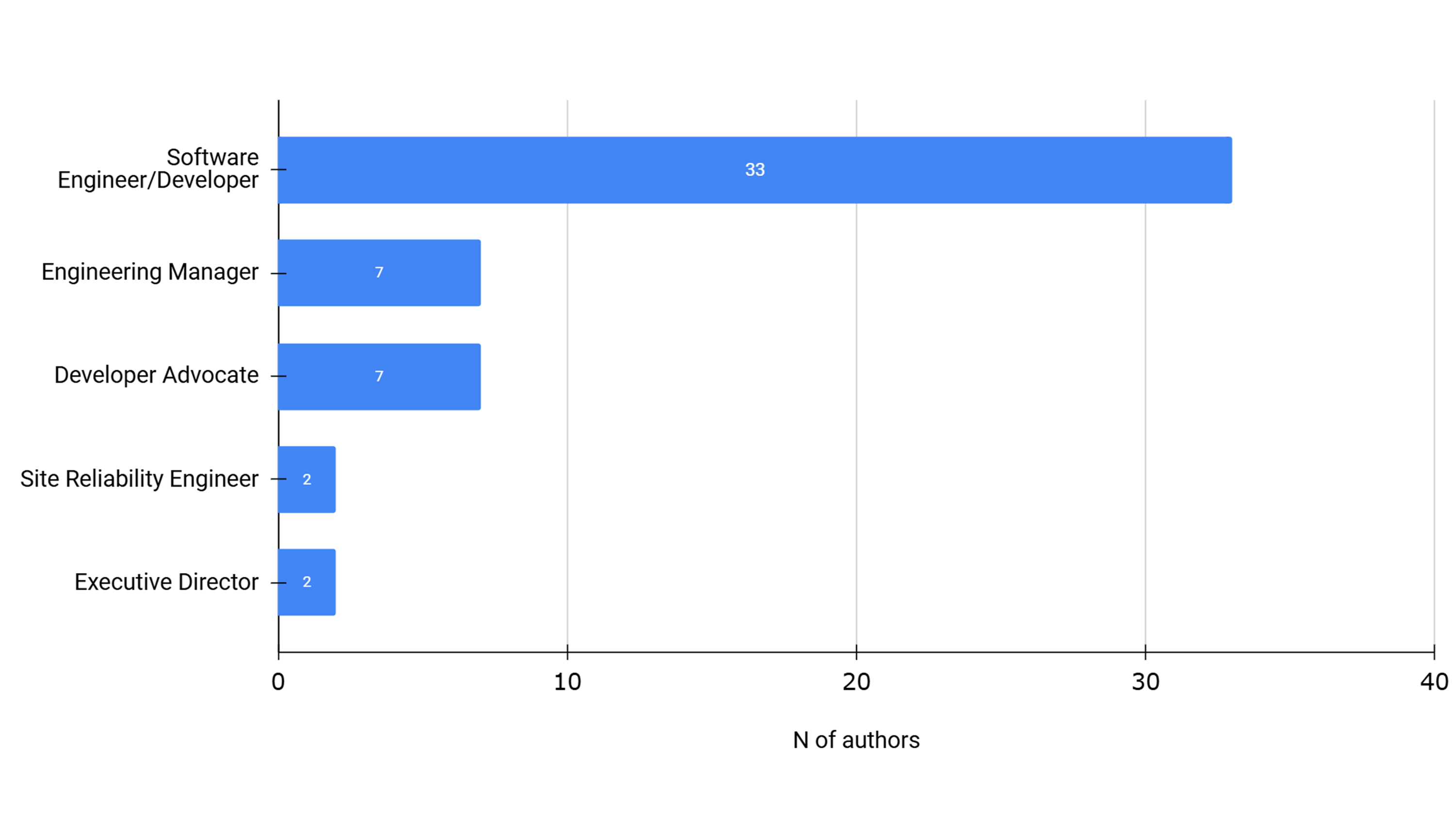}}
\caption{Occupation of the authors.}
\Description[Occupation of the authors.]{}
\label{fig:occupation_chart}
\end{figure}

\textbf{Compassion} is another facet of empathy mentioned by 14 practitioners. It refers to caring for others, as P48 explained: ``\textit{...caring about the people you work with, not just the work you do, A.K.A fostering empathy.}'' 
Practitioners also mentioned \textbf{emotional sharing} (n = 13). It means sharing the emotions or the emotional state of another person. As P54 cited: ``\textit{[it] refers to the sensations and feelings we get in response to others’ emotions; this can include mirroring what that person is feeling or feeling stressed when we detect another’s fear or anxiety.}''

\textbf{Key finding 1.}  \textit{Understanding} is the most cited meaning of empathy for software practitioners. They also view empathy as \textit{perspective taking}, \textit{embodiment}, \textit{compassion}, and \textit{emotional sharing}.  

\subsection{RQ2 - What are the barriers to empathy in software engineering?}
\label{sec:RQ2}

\begin{figure}[b]
\centerline{\includegraphics[scale=0.42, trim=0cm 4cm 0cm 3cm, clip]{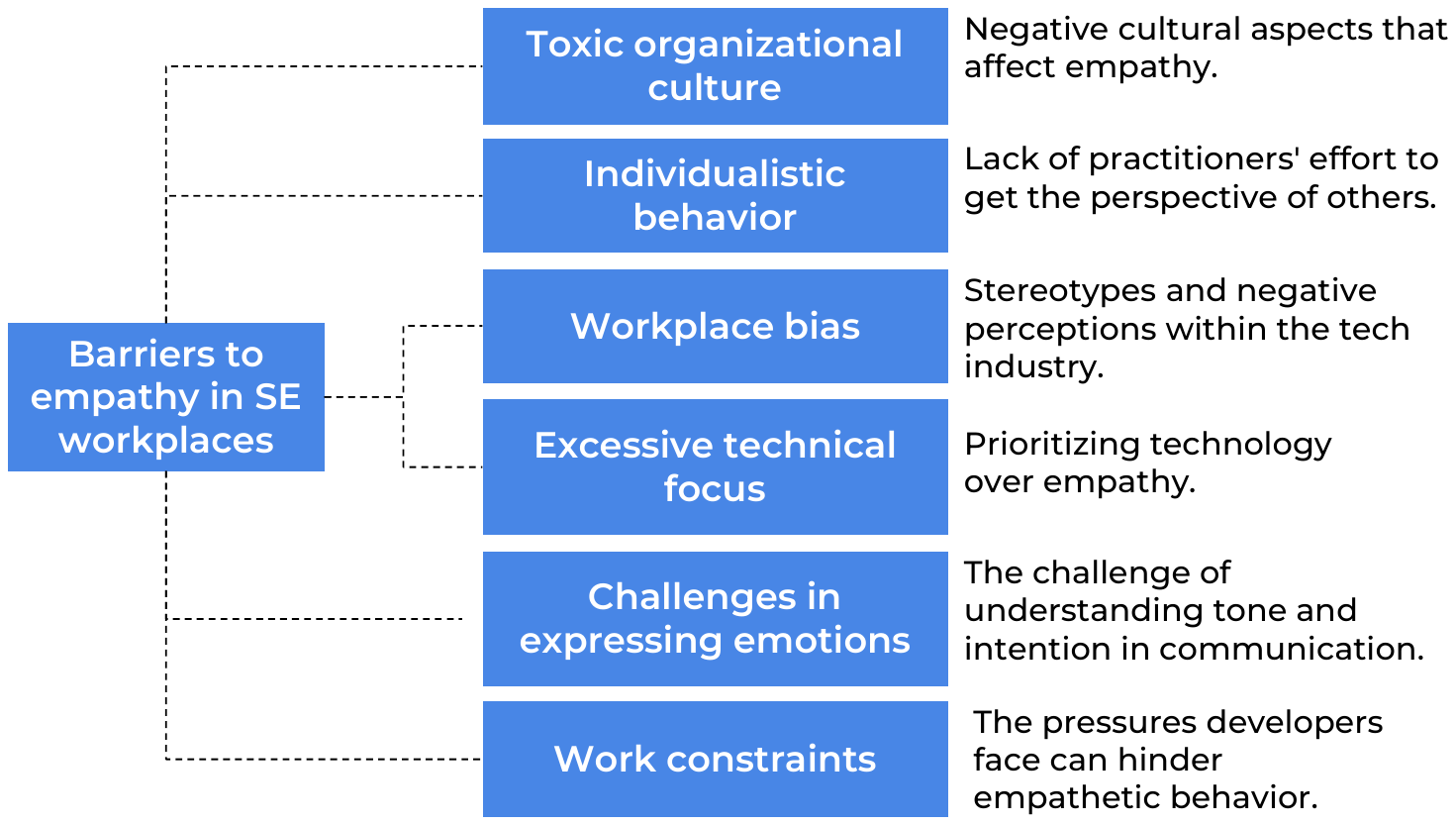}}
\caption{Map of the barriers to empathy according to software practitioners.}
\Description[Barriers to empathy according to software practitioners.]{}
\label{fig:barriers}
\end{figure}

Regarding RQ2, we investigated software practitioners' barriers to practicing empathy in workplace environments. We coded six barriers according to their perspective.
Figure \ref{fig:barriers} presents a map of the barriers found.
The barriers \textit{toxic organizational culture}, \textit{individualistic behavior}, and \textit{workplace bias} were the most commonly found in the analyzed articles. In Appendix \ref{sec:app-codes}, Table \ref{tab:barriers} details the extracted codes. It presents the barriers, their description, their frequency, and the practitioners who mention them.

In the context of this work, \textbf{toxic organizational culture} refers to the negative aspects that affect empathy, including blame culture, dismissive attitudes and comments, lack of support, harmful language, and derogatory terms. It is prevalent in some workplaces, particularly within the tech industry. It can discourage, demean, and isolate individuals and ultimately hinder empathy. It reflects a culture in software engineering of suppressing emotions and empathy in favor of logic, efficiency, and technical prowess. A culture that undervalues empathy in a field traditionally dominated by logic and technical skills. P24 explained how a toxic culture can make it difficult to practice empathy: ``\textit{Practicing that skill can be propelled or held back by the culture our organizations adopt. Practicing empathy can be extremely difficult if we exist in a culture where our humanity is squelched.''} 

\textbf{Individualistic behavior} refers to the attitude of some practitioners who prioritize their methods and preferences over collaboration and support, lacking the effort to get the perspective of others, as P11 detailed: ``\textit{there are 'senior' developers out there that want to do it their way and don't care what others think.}''

\textbf{Workplace bias} refers to stereotypes, such as the ``nerd'' being an unempathetic jerk, perpetuating harmful attitudes and behaviors, suffering, and negative perceptions within the tech industry. It also reflects a tendency to judge or criticize without fully understanding the context, constraints, and challenges that influenced previous developers when encountering legacy codebases. P23 discussed how it hinders empathy in the SE work environment: ``\textit{From the early days of tech, empathy, and emotion have been pushed to the side in favor of cold, hard logic. As software became male-dominated in the 70’s and 80’s, the image of the computer programmer as a socially awkward introvert took shape.}''

The barriers \textbf{excessive technical focus} and \textbf{challenges in expressing emotions} were found in nine and six articles, respectively. The first refers to prioritizing technology, innovation, task completion over empathetic interactions, training for developing empathetic skills, and guidance about practicing empathy. In the following quote, P32 explained it: ``\textit{When you think about software development, the things that usually might come to your mind are coding practices, frameworks, agile workflows, and the newest technology. Many developers forget about the importance of empathizing with the users of our application in the design stage of software development.}'' Whilst \textbf{challenges in expressing emotions} reflects the challenge of accurately understanding tone and intention through written communication without the benefit of vocal inflection, facial expressions, or body language aid in face-to-face conversations. P6 mentioned the challenges in expressing emotions during code reviews: ``\textit{it is difficult to communicate tone properly through text... we all interpret comments in our own way; the impact the reader gains can be completely different than what the writer's intent.}''

Lastly, \textbf{work constraints} were mentioned in two articles. It refers to the pressures developers face, such as tight deadlines and the need to quickly deliver functional code, which can impede the ability to code with empathy and consideration for maintainability, as P55 detailed: ``\textit{several factors are working against even the most experienced developer. You might be under a tight deadline or building out a proof of concept... writing the most elegant and well-organized code possible might take a back seat to just making it work and getting the project out the door.}''

\textbf{Key finding 2.}  Practitioners mention six barriers to sustaining empathy in their workplace: \textit{toxic organizational culture}, \textit{individualistic behavior}, \textit{workplace bias}, \textit{excessive technical focus}, \textit{challenges in expressing emotions}, and \textit{work constraints}.

\subsection{RQ3 - How to practice empathy in software engineering?}

Considering the RQ3, we examined how software workers exercised empathy in their professional practice. We coded 15 empathetic practices according to them. Table \ref{tab:practices} presents the practices, their description, their frequency, and the practitioners who mention them. The practices of \textit{understanding the stakeholders},  \textit{considering different perspectives}, and \textit{adopting good programming practices} were most frequently mentioned in the analyzed articles. For concision, we describe in depth the five most cited practices and present excerpts from the articles to exemplify them. The remaining practices are described in Appendix \ref{sec:app-codes}, Table \ref{tab:practices}. Besides, the entire data set is available in the experimental package \cite{cerqueira_dataset_2025}.

Regarding the context of this study, \textbf{understanding the stakeholders} means understanding what the stakeholders think or feel (including users, clients, customers, and other team members). For instance, P30 mentions how understanding other colleagues can have a positive impact on collaboration and the work environment: \textit{``understanding of the people you work with will make you a much more pleasant colleague and will aid in collaboration and making your work environment a constructive and effective place to be.''}

\textbf{Considering different perspectives}	refers to taking other stakeholders' perspectives, as stated by P23:
\textit{``empathy helps improve the code itself. If a developer has been trained to view their code through other people's eyes, they’ll avoid putting devils in the details."} and P29: \textit{``If we do not see the world from other perspectives, we cannot gather feedback to build more engaging software and products.''}

\textbf{Adopting good programming practices} refers to thinking about future code maintainers and adopting good practices for coding. Practitioners mention choosing good variable names, simple code, and design, maintaining a consistent architecture, adopting single responsibility, fixing what is possible,  using abstractions, modularization, documenting, and adding comments to the code. For instance, P43 discusses the empathy of maintainable code: \textit{``making internal code maintainable is an empathetic behavior\dots''} It reflects care for future developers who will read, maintain, or extend the code. By writing clear, consistent, and maintainable code, developers demonstrate consideration for others' time, effort, and cognitive load.

\textbf{Being compassionate} means being polite, kind, and compassionate to stakeholders. P45 discusses how to be compassionate during the code review:
\textit{``You need to be kind when you comment on the pull request without being rude.''}

\textbf{Running tests} refers to testing the software as mentioned by P43: 
\textit{``Providing automated tests\dots is empathetic because it involves thinking beyond oneself. It is a pattern of `paying it forward'\dots It will allow your team to thrive and build a culture of empathy.''}. It involves anticipating the needs of future maintainers who will interact with the codebase in the future. Besides, it entails preventing users from suffering because of errors present in the code.

\textbf{Actively listening} means actively listening to the needs of the stakeholders and asking meaningful questions, as stated by P42: \textit{``Active listening can greatly enhance your ability to empathize\dots Actively listen to your colleagues, showing genuine interest in their thoughts and feelings.''}

\textbf{Key finding 3.} Practitioners have mentioned 15 practices to include empathy in their activities. The most cited were \textit{understanding the stakeholders} and \textit{considering different perspectives}.

\subsection{RQ4: What are the effects of practicing empathy in software engineering?}

Finally, to answer RQ4, we coded practitioners' perceptions about the effects of practicing empathy in SE.  We identified 30 effects of practicing empathy based on the practitioners' perceptions. They reported that empathy can enhance code quality and further improve communication, trust, and collaboration among stakeholders, which were the most frequently mentioned effects. Table \ref{tab:empathy_effects}, in Appendix \ref{sec:app-codes}, summarizes the codes of this category. It presents the effects, description, frequency, and the practitioners who mention them.  

Practicing empathy results in better \textbf{software quality} and products that truly meet the users' needs by considering their experiences and challenges. In the following quote, P36 describes how empathy can affect the software quality: \textit{``The quality of code can\dots be in direct relationship to the team empathy for fellow engineers who follow in their footsteps\dots an engineer who consistently writes good quality code\dots care for those who will one day inherit her code base.''}

Empathy can enhance \textbf{communication} between the stakeholders. It helps mitigate conflicts and promotes a more supportive work environment. For instance, P30 mentions how empathy can improve communication: \textit{``Practicing empathy, walking that proverbial mile in our colleagues’ shoes, will make reaching solutions easier, make communication between disciplines less tense}.''

For \textbf{trust}, P44 mentions establishing trust among team members by practicing empathy: \textit{``Build a sense of trust and ease with them so they feel comfortable asking you for help.''} Practitioners also reported that empathetic interactions improve team dynamics and a \textbf{better work} culture, as members feel valued and understood, boosting morale, increasing motivation, and raising \textbf{employee retention}. For example, P50 mentions how empathy can improve the \textit{work culture} and \textit{employee retention}: 
\textit{``This type of engineering cultivates a supportive and healthy culture and in turn will directly affect the business by nurturing and retaining top talent.''}

Additionally, they reported how empathy within SE teams can significantly impact levels of \textbf{anxiety}, \textbf{stress}, \textbf{burnout}, and \textbf{frustration} among practitioners. Besides empathetic communication, kindness, and compassion can help prevent frustration or burnout. P48 explains how practicing empathy builds a more supportive environment that can reduce workers' anxiety and stress: \textit{``an empathic workplace is... one with less work-related anxiety/stress, one with a better output quality and a better employee retention.''}

\textbf{Key finding 4.} Software practitioners have discussed the effects of empathetic practices in their projects. Empathy can mainly improve \textit{software quality}, \textit{communication}, \textit{trust}, \textit{collaboration}, and \textit{work culture}, and reduce \textit{anxiety}, \textit{stress}, \textit{burnout}, and \textit{frustration}.

\subsection{A Conceptual Framework of Empathy in Software Engineering}\label{sec:framework}

\begin{landscape}
\begin{figure}[p]
  \centering
  \includegraphics[scale=0.76]{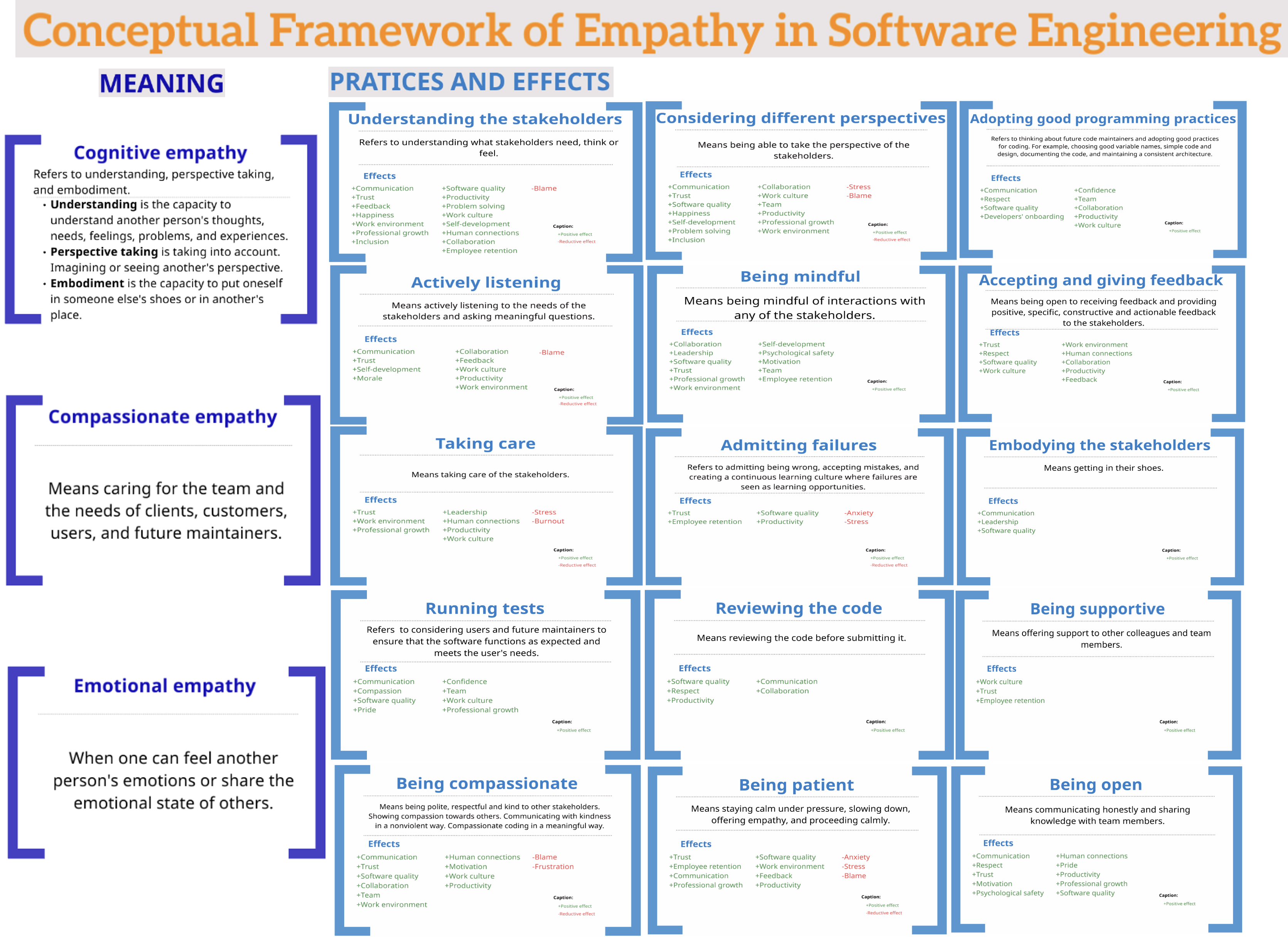}
  \caption{High-level View of the Conceptual Framework of Empathy in SE. A high-resolution and interactive version is \href{https://prezi.com/view/lo0tUsRa7YjJ0BsY2aSN/}{available online} for further exploration.}
  \label{fig:framework}
  \Description[Conceptual Framework of Empathy in SE.]{}
\end{figure}
\end{landscape}

After synthesizing our findings, we propose a framework to systematize the body of knowledge on empathy in SE. In the previous work \cite{cerqueira2023sbes}, we proposed a preliminary version of the conceptual framework based on our findings from the DEV community. However, as we continued with the analysis, we now have a more comprehensive database due to our data analysis from an additional source, the Medium platform, allowing us to refine the framework. Figure \ref{fig:framework} presents the framework overview. 

\begin{figure}[ht]
\centerline{\includegraphics[scale=0.4]{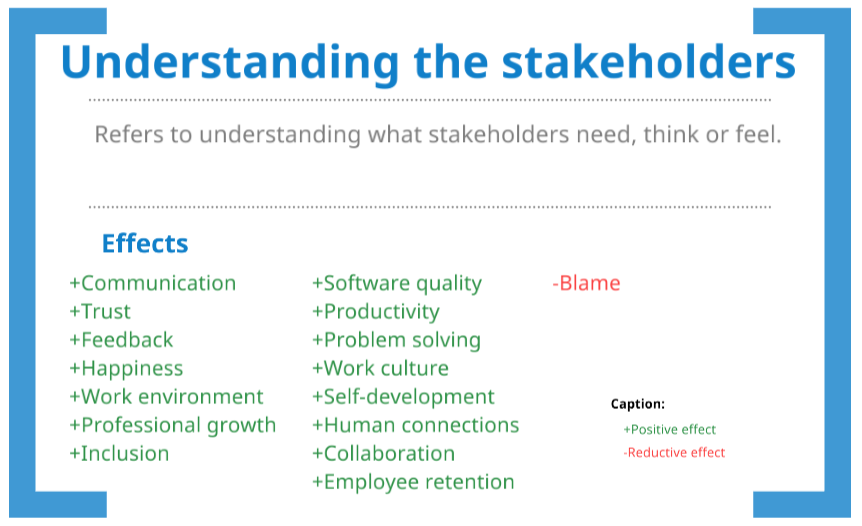}}
 \caption{Low-level View of the Conceptual Framework of Empathy in SE.}
 \label{fig:framework-low}
  \Description[Low-level view Conceptual Framework of Empathy in SE.]{}
\end{figure}

Based on the work of \citet{rocha2021conceptual}, we used the Prezi tool to create a framework and publish it online \footnote{The complete framework is available at \url{https://prezi.com/view/lo0tUsRa7YjJ0BsY2aSN/} and in the replication package}. The framework comprises two navigation levels: a high level representing the overview of the framework and a low level detailing the meaning of empathy, the practices, and their effects. The navigation levels allow interaction with the framework. From the framework overview, it is possible to choose an element to visualize its details. For example, Figure \ref{fig:framework-low} shows the practice \textit{understanding the stakeholders} and its effects. The low-level view contains the details of all elements. As discussed in section \ref{relating_codes}, each practice presents its related effects. They are represented in green or red, with a plus or minus signal if it's positive or reductive, respectively. For instance, the practice \textit{understanding the stakeholders} can positively affect \textit{communication, trust, feedback, happiness, the work environment, professional growth, inclusion, software quality, productivity, problem-solving, work culture, self-development, human connections, collaboration}, and \textit{employee retention}. Besides, it can also reduce \textit{blame}. The framework guide explains the concepts included in the framework, how to navigate it, and its elements (see Figure \ref{fig:framework-guide}) to assist in exploring it.

Following, we discuss how software practitioners can apply empathy-driven practices from our framework in real SE environments: 

\textbf{Onboarding:} For instance, a senior engineer welcomes a new colleague by scheduling regular check-ins, shares curated materials, and simplifies explanations when the newcomer struggles with domain-specific terminology. It involves \textit{actively listening} to newcomers' questions and challenges, showing patience and openness to different learning curves. \textit{Being supportive} by offering help proactively and sharing relevant documentation, and \textit{adopting good programming practices} to reduce the cognitive load for new team members. 

\textbf{Coding:} For example, a developer refactors a function to be more readable and adds clear comments, making the code easier to understand for future contributors. It involves \textit{taking care of the stakeholders} by considering how future maintainers or end users will interact with the system, \textit{being mindful} of complexity, accessibility, and maintainability when writing code, and \textit{admitting failures} by documenting limitations or technical debt transparently.

\textbf{Code Review:} Instead of rejecting a non-standard solution outright, a reviewer asks clarifying questions and commends the effort, then discusses alternatives that better align with system conventions. It encompasses \textit{considering different perspectives} when evaluating a peer's code, recognizing that different implementation approaches may stem from valid reasoning,  \textit{being compassionate} by offering constructive, respectful feedback that focuses on improvement rather than blame and \textit{Understanding the stakeholders} by keeping the review’s impact on the developer's confidence in mind.

\color{black}

%% file: content/5_discussion.tex
This section summarizes how we evolved the empirical knowledge on empathy in SE. First, it discusses the results, focusing on each RQ and connecting them to related work. Then, it compares the findings from Medium and DEV communities. Lastly, it discusses how our conceptual framework of empathy in SE can support software workers in practicing empathy, overcoming the barriers to empathy, and developing empathetic skills. 

\begin{figure}[b]
\centerline{\includegraphics[width=\linewidth]{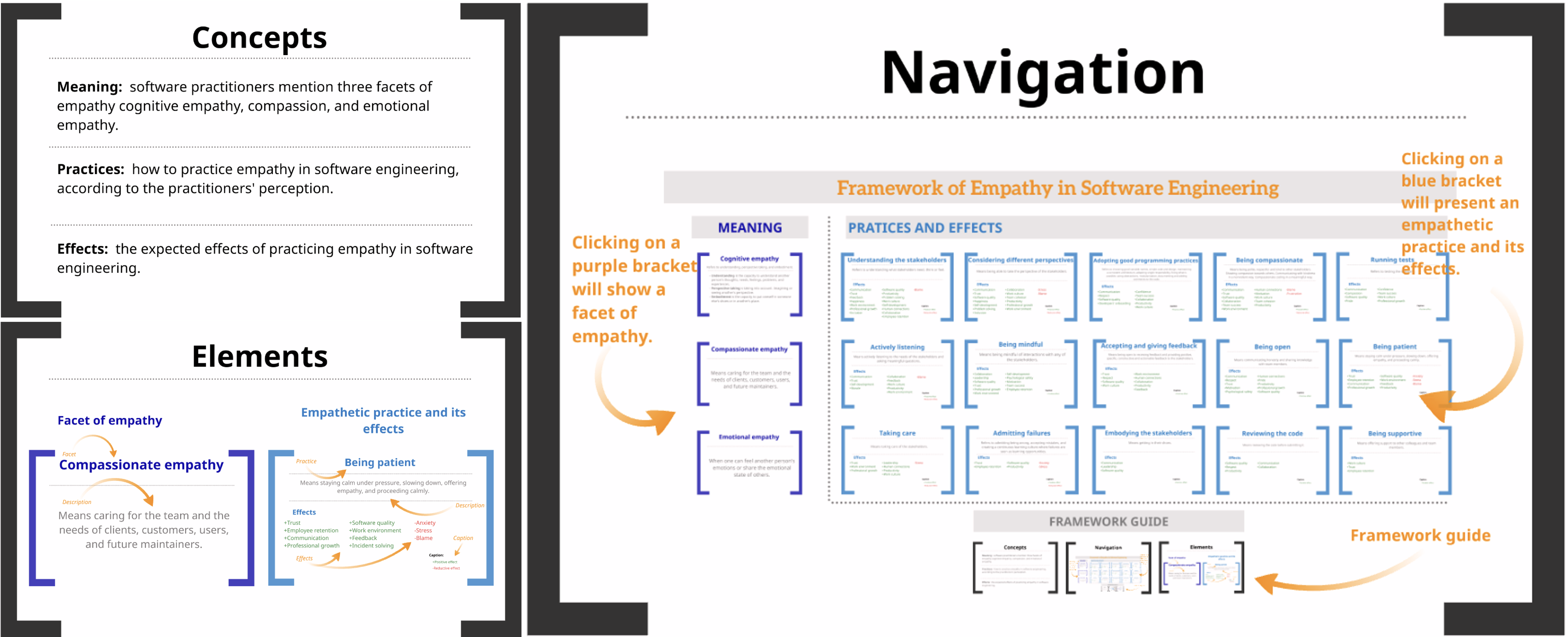}}
\caption{Conceptual Framework Guide.}
\label{fig:framework-guide}
\Description[Conceptual Framework Guide.]{}
\end{figure}

\subsection{Answering the Research Questions}

\subsubsection{RQ1 - What does empathy mean for software practitioners?}

Empathy in SE has five meanings: \textit{understanding, perspective-taking, embodiment, compassion, and emotional sharing}. Practitioners have mentioned these concepts aligning with three facets of empathy: cognitive empathy, compassionate empathy, and emotional empathy \cite{decety2020empathy}. Understanding, as the capacity to understand another person's thoughts, needs, feelings, problems, and experiences, is the most cited theme, followed by compassion and embodiment.

Table~\ref{tab:comparison} compares the empathy meanings found in our study with what is reported in related work (RQ1). The themes \textit{understanding the stakeholders, perspective taking} and \textit{embodiment} were aligned with the engineers' understanding of empathy \cite{hess2016voices}. 
Furthermore, by comparing these themes with the empathy facets proposed by~\citet{decety2021emergence}, we notice that they cover the cognitive aspect of empathy, making it the most relevant for practitioners. However, our themes did not comprise the emotional regulation facet ~\cite{decety2021emergence}. 
The engineering model of \citet{walther2017model} also includes \textit{perspective taking}. Besides, they also cite \textit{affective sharing}, which relates to \textit{emotional sharing}. 
\citet{devathasan2024deciphering} and \citet{gunatilake2023} cited five dimensions of empathy based on the PAM model: emotion contagion, sympathy, helping behavior, cognitive empathy, and empathetic accuracy ~\cite{preston2007perception}. Both emphasize the multi-faceted nature of empathy, including cognitive aspects like \textit{perspective-taking} and \textit{understanding}, besides \textit{compassion} and emotional sharing.
The dimensions of empathy outlined in the PAM align closely with our findings in SE.  PAM integrates empathy as a shared emotional experience rooted in motor and emotional behaviors. On the other hand, our findings include specificity by emphasizing \textit{embodiment} and \textit{understanding}. These parallels suggest that PAM’s broad view of empathy can effectively encompass our SE-specific categories, while the slight distinctions reflect nuances in empathy’s expression within SE contexts.

Yet, these models \cite{hess2016voices, walther2017model, preston2007perception} do not mention the compassion facet of empathy identified in our work.  
\color{black}
Our findings regarding the meaning of empathy help to comprehend this human aspect from the perspective of software practitioners. Besides, we lay the groundwork to define this concept within SE. Our results help strengthen its comprehension and address the lack of consensus discussed in Section \ref{sec:related_work}.

\color{blue}

\begin{table}[t]
\small
\caption{The meaning of empathy for software practitioners: comparison with related work.}
\label{tab:comparison}
\begin{tabular}{p{2.5cm}p{2.8cm}p{3cm}p{2.8cm}p{2.3cm}}
\toprule
\textbf{Software Practitioners (our work)} & \textbf{Empathy framework \cite{decety2021emergence}} & \textbf{Engineers’ understanding \cite{hess2016voices}} & \textbf{Engineering Model \cite{walther2017model}} & \textbf{PAM \cite{preston2007perception}} \\
\midrule
Perspective taking & Cognitive empathy & Perspective taking & Perspective taking & Cognitive empathy \\
Embodiment & Cognitive empathy & Embodiment & - & - \\
Understanding & Cognitive empathy & Understanding & Self and Other awareness & Empathetic accuracy \\
Emotional sharing & Emotional empathy & - & Affective sharing & Emotion contagion \\
Compassion & Compassion & - & - & Sympathy, helping behavior \\
- & Emotional regulation & - & Emotion regulation & - \\
- & - & Connectedness & - & - \\
- & - & - & Mode switching & - \\
- & - & - & - & Conscious helping behavior \\
\bottomrule
\end{tabular}
\end{table}

\color{black}

\subsubsection{RQ2 - What are the barriers to empathy in software engineering?}

We coded six barriers to empathy faced by software practitioners. They are presented in Table \ref{tab:barriers}. 
The related work identified several barriers to empathy across professions, such as high workload, lack of organizational support, burnout, and time constraints in healthcare, which reduce empathy by leaving professionals emotionally drained and with limited time to connect with others \cite{shanafelt2002burnout, hojat2009devil, wilkinson2017examining}. Studies also highlighted specific challenges in SE, such as poor connection with users, unfamiliarity, and a strong focus on task-oriented rather than relational interactions \cite{gunatilake2024enablers}. Similarly, our findings reinforce these themes, identifying barriers such as \textit{focusing on technical aspects}, having an \textit{individualistic behavior}, building a \textit{toxic organizational culture}, \textit{workplace bias}, \textit{challenges in expressing emotions}, and \textit{working constraints}. They also extend this understanding by uncovering additional barriers beyond user interaction. Our findings uniquely emphasize how SE-specific factors like a technical focus and communication challenges in digital interactions, suggesting that these barriers are often contextual and influenced by industry-specific practices.

\subsubsection{RQ3- How to practice empathy in software engineering?}

Based on the premise that action is essential for professionals when incorporating empathy in their practice \cite{gerdes2009social}, we identified 15 empathetic practices mentioned by software practitioners. Table \ref{tab:practices} presents these practices. They offer an instrument for practitioners to understand how to exercise in their workplace. They also advance the topic with new insights into practitioners' dynamics and their impact on SE environments.

These practices closely reflect the qualities of empathic engineering described by \citet{hess2016voices}. \textit{Understanding stakeholders} and \textit{considering different perspectives} show how practitioners actively seek to understand others’ needs and integrate that knowledge into their work. \textit{Embodying the stakeholders} aligns with adopting their perspectives, guiding engineers to make socially informed decisions. \textit{Running tests}, \textit{reviewing code}, and \textit{adopting good programming practices} ensure that engineering decisions are of the highest quality, considering their broader impact.
They are technical responsibilities, but also expressions of empathy. These actions demonstrate care for teammates and users by improving code quality, reducing future errors, and making the development process smoother for others. By writing clear code and reliable tests, developers help teammates (and their future selves) understand whether changes break existing functionality. This reduces cognitive load, increases trust, and contributes to a more supportive and collaborative environment. For instance, running and maintaining automated tests helps prevent bugs and ensures stability, which shows consideration for those who will maintain or build upon the code and the users. Similarly, thoughtful code reviews provide constructive feedback and support learning, while clean, well-structured code reflects a desire to reduce friction for collaborators. Beyond technical best practices, they are also a behavior that supports team well-being and effectiveness, aligning with an empathic culture. 

\textit{Accepting and giving feedback}, \textit{being patient}, and \textit{actively listening} demonstrate a conscious effort to adjust behavior according to others' needs, fostering an open and supportive environment. \textit{Being compassionate}, \textit{taking care}, \textit{being mindful} reflect a caring approach, enabling engineers to create a work culture that values empathy. Finally, \textit{admitting failures} and \textit{being open} emphasize humility and transparency, strengthening connections between stakeholders. Together, these practices reveal how empathy in SE goes beyond technical considerations, ensuring practitioners remain attentive to human elements in their work.

\subsubsection{RQ4 - What are the effects of practicing empathy in software engineering?}
Our prior research was the first work to address the effects of empathy in SE \cite{cerqueira2023sbes} and also to focus on its impact on practitioners’ well-being and mental health \cite{cerqueira2024empathy}. In this study, we extend our previous findings by listing 30 different effects of practicing empathy in SE. They are summarized in Table \ref{tab:empathy_effects}. Listing these effects is crucial to understanding the importance of fostering empathy in SE work environments.

The description of empathy’s effects cited by \citet{hess2016voices} closely aligns with our findings on the positive impacts of empathy in SE. For instance, they mentioned improved solution quality, interpersonal relations, leadership skills, and motivation to help others. According to our findings, practicing empathy enhances software quality by creating user-centered products with fewer bugs, reflecting more effective and high-quality solutions. Improved leadership, communication, collaboration, and trust among team members, clients, and stakeholders parallels \citet{hess2016voices}'s outcomes of stronger interpersonal relations and better management skills. The professional growth and self-development associated with empathy promote career advancement and effective leadership, giving engineers the confidence and interpersonal skills they need for career progression. \citet{hess2016voices} notion of empathy-driven motivation to help others is mirrored in how empathy in SE builds strong human connections and improves customer satisfaction and employee retention.

\subsection{Comparing the findings from Medium and DEV}

In this section, we discuss our results by comparing our findings in each community.

The RQ1 was addressed in our previous work \cite{cerqueira2023sbes, cerqueira2024empathy}. However, upon revisiting it, we were able to reinforce our findings with the perspectives of professionals from both data sources. Comparing the data from the DEV community and Medium, we have found the themes in both data sets, which can strengthen the validity and reliability of the research process and findings. The consistency of the findings suggests that the identified themes are not unique to a single data source but are potentially generalizable and reflective of broader patterns.

The RQ2 was not included in our previous study. We revisited the texts from the DEV community to address this question. We found the barriers mentioned by practitioners in both communities. Regarding RQ3, in our previous work \cite{cerqueira2023sbes}, we coded 19 empathetic practices. However, as we explain in Section \ref{reviewing}, we merged some practices to better reflect their definition. When analyzing the data from DEV and Medium, the practitioners recommend the coded practices on both platforms.

Concerning RQ4, when comparing the results to our previous work \cite{cerqueira2023sbes}, we found 28 effects of practicing empathy in the DEV community. 
Practitioners mentioned some effects just in the DEV community. For instance, the positive impact on compassion, developers' onboarding, morale, and pride. One DEV practitioner also mentioned how empathy can prevent burnout. On the other hand, we found two additional effects when analyzing the articles from the Medium platform. They noted that empathy can improve software practitioners' confidence and reduce anxiety. Thus, we coded 30 effects of practicing empathy when analyzing the two communities.

Our analysis of Medium and DEV articles found consistent results across RQ1, RQ2, and RQ3, reinforcing the reliability of our findings on empathy's meaning, barriers, and practices. However, RQ4 results showed slight differences between the two platforms, suggesting that diverse sources can reveal nuanced variations in practitioners' perspectives. This highlights the importance of investigating multiple data sources to understand the topic thoroughly.

%% file: content/5.1_framework.tex
\subsection{How the framework can manage barriers to empathy}

In this section, we discuss how the practices presented in the conceptual framework can help software practitioners address and manage the barriers discussed in Section \ref{sec:RQ2}. 

Practitioners cite a \textbf{toxic work culture} as a significant challenge to empathy in the workplace. It includes blame-shifting, dismissive attitudes, and competition, discouraging empathy. The framework builds trust and motivation through practices like taking care (including regular check-ins), offering support, and accepting and giving constructive feedback, as P34 explained: ``\textit{Often there will be low trust and a high degree of blame across teams, where individuals are punished \dots Leaders who ´walk the floor' build more trust and credibility\dots Leaders need to understand all of the `Empathy for Individuals' section above and act accordingly\dots Leaders who do will build more psychological safety.''}

Practitioners often cultivate an \textbf{individualistic behavior}, focusing on their methods, preferences, or objectives, neglecting collaboration and feedback. The framework suggests that practitioners seek to understand others' perspectives, actively listen, and be more compassionate and mindful of the stakeholders. As P26 mentioned, for example:
\textit{``By avoiding personal attacks and criticism directed at the author, we can instead focus on improving the code itself.''} 

To overcome an \textbf{excessive technical focus}, the framework proposes human-centered practices, encouraging practitioners to balance technical goals with an \textit{understanding of the needs of others}, \textit{considering different perspectives}, and \textit{being mindful} of the stakeholder's needs, for example. As P3 explained \textit{``most of the schools have an excessive focus on technical subjects making some of the future professionals with almost zero people skills as they are leaving the more `human' topics aside''}, highlighting that software professionals must develop their skills beyond technical proficiency: \textit{``Understand the client is not difficult, we just need to empathize with them\dots if you don't understand the needs of your client you most likely don't understand your colleagues either.''} With the practices defined in the framework, we propose a paradigm shift, moving the focus from technology to a human-centered approach based on empathy-guided practices, as P54 stated: \textit{``A human-centered process is a multidisciplinary approach rooted in empathy. To create better products that are better for us, we must empathize with the needs of everyone involved.''}

Another barrier cited by the practitioners is the \textbf{workplace bias}, endorsing stereotypes and viewing specific roles as less empathetic or perpetuating discriminatory views, which can undermine empathy. Our framework encourages diverse viewpoints and promotes inclusion, ensuring that empathy can be more consciously practiced across various situations. P4 explained \textit{``your subconscious mind may not let you experience that person's actual state or condition\dots That's where Empathy plays a very important role. If you are a fellow developer or a lead and practice being empathetic, you can create a better work environment for that person and also help him/her grow eventually}''. According to P4, practicing empathy by \textit{being mindful} of interactions can help overcome subconscious bias in the workplace: \textit{``So being mindful about your daily interactions, be it with anyone in your office or personal life, can help you learn and practice this.''} Besides, to overcome this barrier, software organizations can adopt the framework to promote regular bias-awareness training and activities like perspective-taking exercises. 

Practitioners also mention the \textbf{challenges in expressing emotions} as one of the barriers to empathy. To overtake this, the framework proposes practices like \textit{being compassionate} and \textit{understand the stakeholders}, as P6 described: \textit{``You can only guess, and understanding that you do not know where they are coming from when they make such comments is the first step to empathy. We can show compassion towards the commenter and try to understand their feelings and needs.''} According to P6, following these practices, it is possible to overtake this barrier: \textit{``creating real connections, and being able to communicate with one another on the most human level''}.

\textbf{Work Constraints} such as tight deadlines and pressures to deliver functional code often override empathy considerations for the stakeholders. Introducing empathy-driven practices like \textit{adopting good programming practices}, \textit{running tests}, \textit{embodying the stakeholders}, and \textit{considering different perspectives} can foster empathy, even under tight schedules. For instance, P55 explained that even when working on a tight deadline, practitioners should keep others in mind to overcome this challenge: \textit{``If you use empathy driven development, where you’re thinking about the next person working on a thing, you’re going to make decisions keeping them in mind \dots you can start to employ clean code practices\dots you’ll produce code that doesn’t make life difficult for future developers working on your project.''}

%% file: content/5.2_evaluating_the_framework.tex
\subsection{Empathy experts' assessment of the framework}

We received seven answers\footnote{The raw data is available at our replication package.} representing a $\sim$14\% response rate. We gathered responses from empathy experts across multiple countries, including the United States (3), New Zealand (1), India (1), Cambodia (1), and the Czech Republic (1). Participants held diverse roles, such as Software Developers/Engineers (3), UX/UI Designers (1), Senior Managers/Executives (2), and a CTO/Co-Founder (1). Participants' professional experience ranged from 5 to over 15 years, with most having extensive industry backgrounds. While one UX/UI Designer had 5-10 years of experience and one Software Developer/Engineer had 10-15 years, the majority (four participants) had more than 15 years in the field, including the Senior Managers/Executives and Software Developers/Engineers. Their expertise on empathy varied, with one reporting intermediate knowledge, four identifying as advanced, and two as experts. This distribution highlights a participant pool with substantial industry experience and a strong understanding of empathy in SE.

\subsubsection{Framework influence} Responses to Q5 revealed how the framework changed their perceptions of empathy in SE. Some participants appreciated the framework’s actionable practices, valuing concrete steps they could implement. For instance, a participant mentioned: ``\textit{I like specific practices because they are actionable, bite-sized, to try.}'' Others recognized the importance of coding for future maintainers, emphasizing the need to consider how their code impacts others. While some respondents reported that their perception did not change, this was often due to their long-standing engagement with empathy through leadership, teaching, or extensive experience in the field: ``\textit{I can’t think of anything that has changed. This has been a topic I’ve been diving deeply into for over a decade.''} Additionally, a participant stated, ``\textit{I think at our org we follow these instinctively, it's very much embedded in our culture. Teams where this was fostered unsurprisingly did a lot better},'' acknowledging empathy’s impact on work outcomes and noting that the framework aligned with their existing company culture, reinforcing practices they already followed.

Participants' responses to Q6 highlighted what they learned from the framework. Some appreciated the structured breakdown of empathy into cognitive, compassionate, and emotional facets, finding it helpful for understanding its application in SE: ``\textit{Separating into cognitive, compassionate, and emotional helps segment ways to apply empathy into a structured format}.'' Others valued the actionable empathy practices presented, noting their practicality: ``\textit{There are different facets to empathy, and many easy-to-practice action items}.'' A respondent stated ``\textit{One point that made me think was 'Taking Care'—it has a reductive effect on cashflow stress!}'' They reflected on how empathy influences customer experience and business outcomes, particularly in reducing financial stress. Additionally, a participant recognized empathy’s role in fostering collaboration, especially in remote settings: ``\textit{When there's empathy, people are more open to collaborate \dots We are 'forced' to understand what the person is going through}.'' However, two participants reported no new insights due to prior familiarity with the topic or because the framework did not introduce novel perspectives for them.

\subsubsection{Framework organization}
Responses to Q7 and Q8 revealed a mix of agreement and critiques. 
Most participants agreed (57.14\%) with the elements of the framework, though 42.86\% expressed disagreement.
When answering Q8, they considered the framework conceptually sound but noted challenges in practical implementation: ``\textit{As a concept, I think it's fine, a different thing would be to implement in reality.}'' Particularly in convincing business leaders of its value without a clear link to financial outcomes: ``\textit{Business leaders won't take efforts related to empathy seriously unless they can draw a direct connection to how it impacts their bottom line.}'' Others did not disagree with the framework but felt its presentation could be improved, describing it as overly complex for a short introduction. One participant appreciated how the framework effectively highlighted empathy in SE, while another suggested theoretical refinements, such as incorporating the emotional regulation facet, elements of self-compassion, and considering potential biases in embodiment.

Similarly, responses to Q9 and Q10 indicated that most participants found the framework clear and well-explained. Although most did not find any concepts confusing (71.43\%), two participants stated that some definitions were unclear (Q9), indicating that the framework's organization was unintuitive and pointed out perceived imbalances in the visual representation and distribution of elements. Despite these concerns, no fundamental disagreements about the framework's concepts were raised.

\subsubsection{Completeness, Accuracy, and Usefulness of the Framework}

Regarding completeness (Q11), ratings ranged from 3 to 10 (mean $\sim$6.57), with one participant rating it very low (3) and another considering it fully complete (10). In response to Q12, they suggested improvements to the framework. One participant emphasized the need for more practical context, such as tying the framework to real-world studies, customer success cases, and concrete coding examples. Another highlighted the importance of improving the framework’s narrative, suggesting a reduction in complexity and a more straightforward storyline: ``\textit{I would recommend dramatically reducing the complexity and developing a narrative.}'' Additionally, another participant recommended explicitly defining the criteria for placing elements into each category. While one participant did not propose specific changes, ``\textit{There are probably a few things none of us have considered, so just leaving room to improve},'' but acknowledged room for refinement.

Accuracy ratings (Q13) followed a similar trend, ranging from 5 to 10 (mean $\sim$7.29), with most participants rating it seven or higher. These results suggest general confidence in the framework, though some participants perceive areas for improvement in its accuracy. Responses to Q14 offered perspectives on it. One participant suggested incorporating a broader, cross-functional perspective, particularly drawing insights from product management. Another questioned whether empathy should be considered a framework, arguing that it influences software development but does not function as a structured framework like Scrum. Some responses focused on refining the framework’s organization and including more practical context. Additionally, suggestions included enhancing descriptions by replacing bulleted lists with more explanatory sentences and adding more useful applications to improve accuracy. One participant did not propose specific changes but recognized potential areas for improvement.

In answer to Q15, most participants ($\sim$85.71\%) believed that the practices captured in the framework would influence their SE activities, indicating a strong potential for the framework to impact their work. One participant, however, disagreed, suggesting that the practices may not align with their current approach or needs in SE. The answers to Q16 offered more insights into the framework’s influence on SE activities. One participant indicated that it would impact their decision-making, collaboration, and feedback processes, influencing how they approach tasks such as code refactoring with future maintainers in mind: ``\textit{how I make decisions, from priority calls to delivering feedback, to accepting feedback, how I interact with other engineers in the codebase and if I take on a large refactor from the perspective of tearing through the code base and changing whatever I want, or if I care who's going to be fixing my mistakes in the morning.}'' 
Another participant noted that they already apply many of the suggested practices in their daily interactions with colleagues ``\textit{They already do as, somehow, I apply them. Mostly interacting with my colleagues}.'' One participant found value in the framework’s ability to address an underexplored aspect of software engineering, emphasizing its potential to expand industry perspectives on empathy: ``\textit{I think most people have a simplified view of empathy in general, and how it can be applied to.}'' However, one participant expressed that the framework lacked specific practices tied to empathy. At the same time, another found certain elements helpful but others too abstract for practical application in industry settings: ``\textit{It's not yes and no for me. Some things I follow implicitly already; the other bits are more abstract and less practical, given the realities/constraints of the industry's operation.}'' Overall, the responses to RQ16 reflected appreciation for the framework’s insights and calls for more concrete, actionable guidance.

\subsubsection{Feasibility of the Framework}

The responses to Q17 and Q18 revealed mixed perspectives on whether the effects of practicing empathy, as outlined in the framework, are feasible to achieve. When answering Q17, most participants (71.43\%) believed that the effects of practicing empathy are possible to achieve through the practices captured in the framework (Q17). However, two participants disagreed, suggesting that some aspects of the framework may require further refinement or additional practices to support empathy in SE fully. Analyzing the answers to Q18, we noticed that participants believed that empathy improves productivity, collaboration, and workplace culture, emphasizing the role of open communication, kindness, and psychological safety: ``\textit{With open communication, true collaboration, and kindness, I do not see how productivity and company culture wouldn't benefit}.'' One participant highlighted the concrete effects on business: ``\textit{I do love how this framework builds an active empathy practice that leads to tangible business outcomes}.'' Others noted that while the framework outlines empathy-driven behaviors well, it needs better organization and a stronger link between empathy, coding practices, and business outcomes. Some respondents pointed out limitations: one mentioned that empathy alone is insufficient for success, as other factors also play a role in achieving business goals. Another critiqued the survey, arguing that it imposes a predetermined perspective rather than fully exploring empathy in software development.

Additionally, one respondent highlighted the need for more concrete activities to help developers apply empathy in practice: ``\textit{There are many developers who are craving more human connection in their work, and having concrete and specific activities on how to apply empathy is something I hear folks say they want more of}.'' Another stressed the importance of role models and a sense of belonging in fostering an empathetic culture. While many saw value in the framework, they also suggested refinements to make it more actionable and applicable in real-world software engineering contexts. As one respondent stated: ``\textit{The organization of this framework has some specific activities that could be incredibly valuable if they were better linked to the developer, code, and business benefits.}''

 Responses to Q19 showed a mix of opinions on whether additional practices should be included in the framework. Two participants found the existing practices sufficient, providing a strong foundation. However, others suggested specific additions:

 \begin{itemize}
     \item Managing cognitive load to reduce stress: one respondent emphasized that organizations should manage cognitive load to prevent stress and burnout, which can impact empathy at scale.

\item Promoting discussions on empathy: another participant suggested fostering open talks about empathy in software development, potentially expanding these conversations to include business analysts and product owners.

\item Distinguishing empathy from sympathy: one participant highlighted the need to clarify the difference between empathy (understanding others' perspectives) and sympathy (an emotional reaction based on assumptions).

\item Adding self-compassion and loving-kindness: another participant recommended including these practices, as they help individuals manage personal distress, which can otherwise hinder empathetic behavior.

\item Concerns about abstraction: one respondent criticized the framework as too academic and detached from real-world software development, arguing that it lacks practical applicability.

 \end{itemize}

These insights suggest that while many find the framework valuable, some feel it could still be expanded to include additional factors.

\subsubsection{Assess the structure and use of the framework}

\begin{figure}[b]
\centerline{\includegraphics[scale=0.6, trim=0cm 1cm 0cm 2cm, clip]{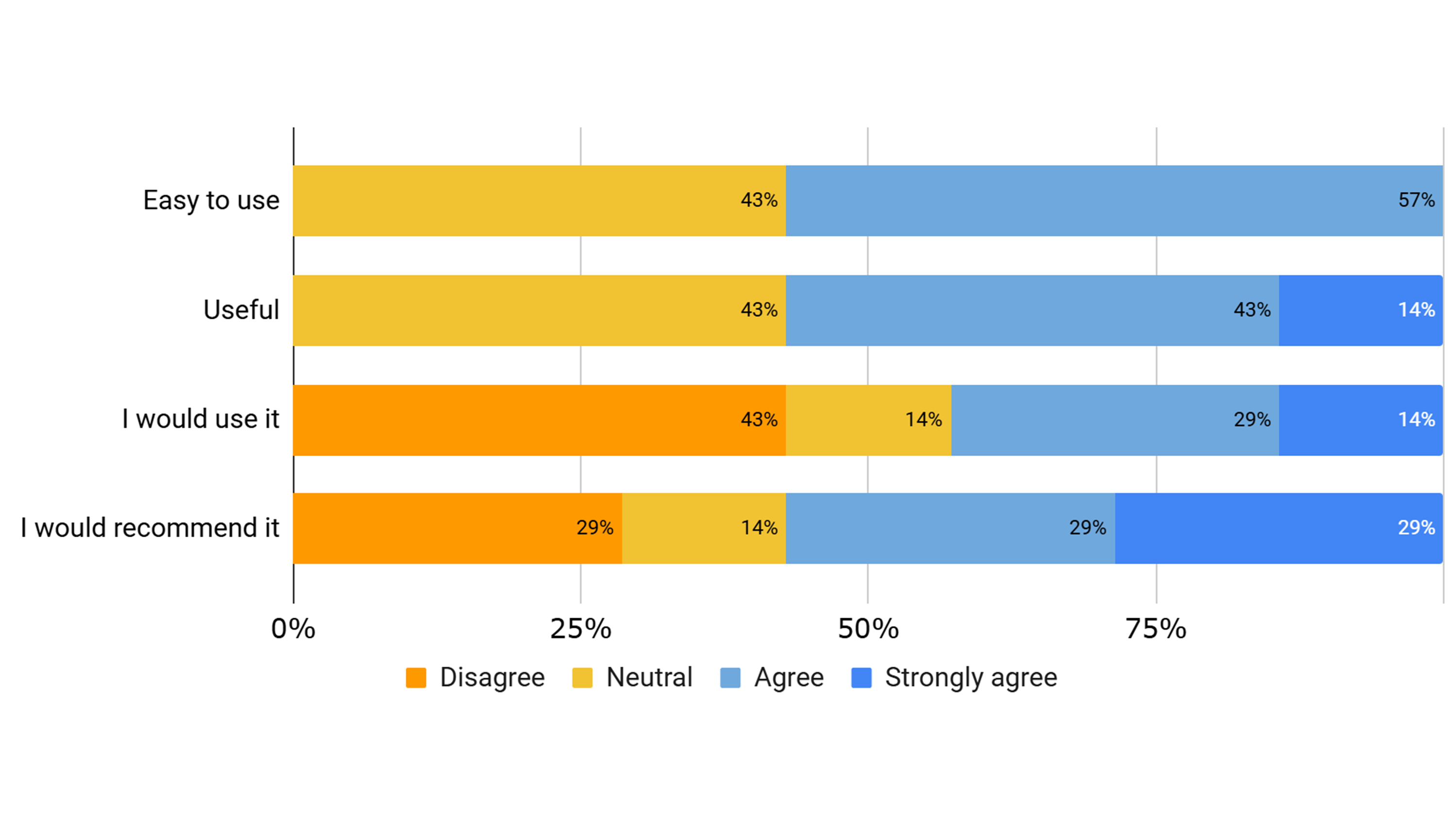}}
\caption{Practitioners' assessment of the framework's usefulness.}
\Description[Practitioners' assessment of the framework.]{}
\label{fig:grafico_survey_use}
\end{figure}

In Q20, participants generally found the helpful framework and easy to use for practicing empathy in software engineering, with around 71.43\% agreeing on both aspects. Figure \ref{fig:grafico_survey_use} summarizes their assessment of the framework's usefulness.
However, opinions on the regular use in the workplace were mixed, with 43\% of participants either disagreeing or remaining neutral about its frequent use, while 57\% predicted they would use it regularly. When it came to recommending the framework to other practitioners, 58\% of participants indicated they would recommend it. However, two participants disagreed, and one remained neutral, suggesting that while the framework is valuable to many, some concerns about its applicability or effectiveness in real-world settings remain. 

In the answers to Q21, participants suggested various ways to use the framework in their work, though some noted that its effectiveness depends on broader team or organizational adoption. For instance, one participant stated ``\textit{I am all for it, but I do not work in isolation. It would need at least a team implementation, if not a whole organization one}.'' Some saw value in using it for training and culture building, ensuring empathy-driven behaviors become part of organizational norms. Others suggested incorporating it into reward systems to incentivize consistent practice. One participant emphasized its applicability in daily interactions, such as meetings, code reviews, and mentoring: ``\textit{In pretty much every area where I need to interact with people: From meetings to code reviews, including mentoring.}'' While another noted its potential for discussing interpersonal skills with engineers. Some respondents viewed the framework as a guiding philosophy for role modeling rather than a rigid structure: ``\textit{I use all this as a north star, something to inculcate and role model in teams I work with.}'' However, one critique was that the framework could be more useful if it were better organized. 

\subsubsection{Framework for Mitigating Empathy-Related Challenges}
Participants highlighted the barriers to empathy they believe can be overcome in their work environments when answering Q22. The most common barriers were excessive technical focus (reported by 71.43\% of participants) and individualistic behaviors (reported by 85.71\%). Other participants reported challenges in expressing emotions, workplace bias, work constraints (each reported by 57.14\% of participants), and toxic organizational culture (reported by 42.86\%). One participant stated that none of these barriers could be overcome in their environment, suggesting complex challenges. These responses indicate that while many barriers to empathy exist, practitioners see opportunities to address them through workplace changes. 

\begin{figure}[b]
\centerline{\includegraphics[width=\linewidth,trim=0cm 3cm 0cm 2.9cm, clip]{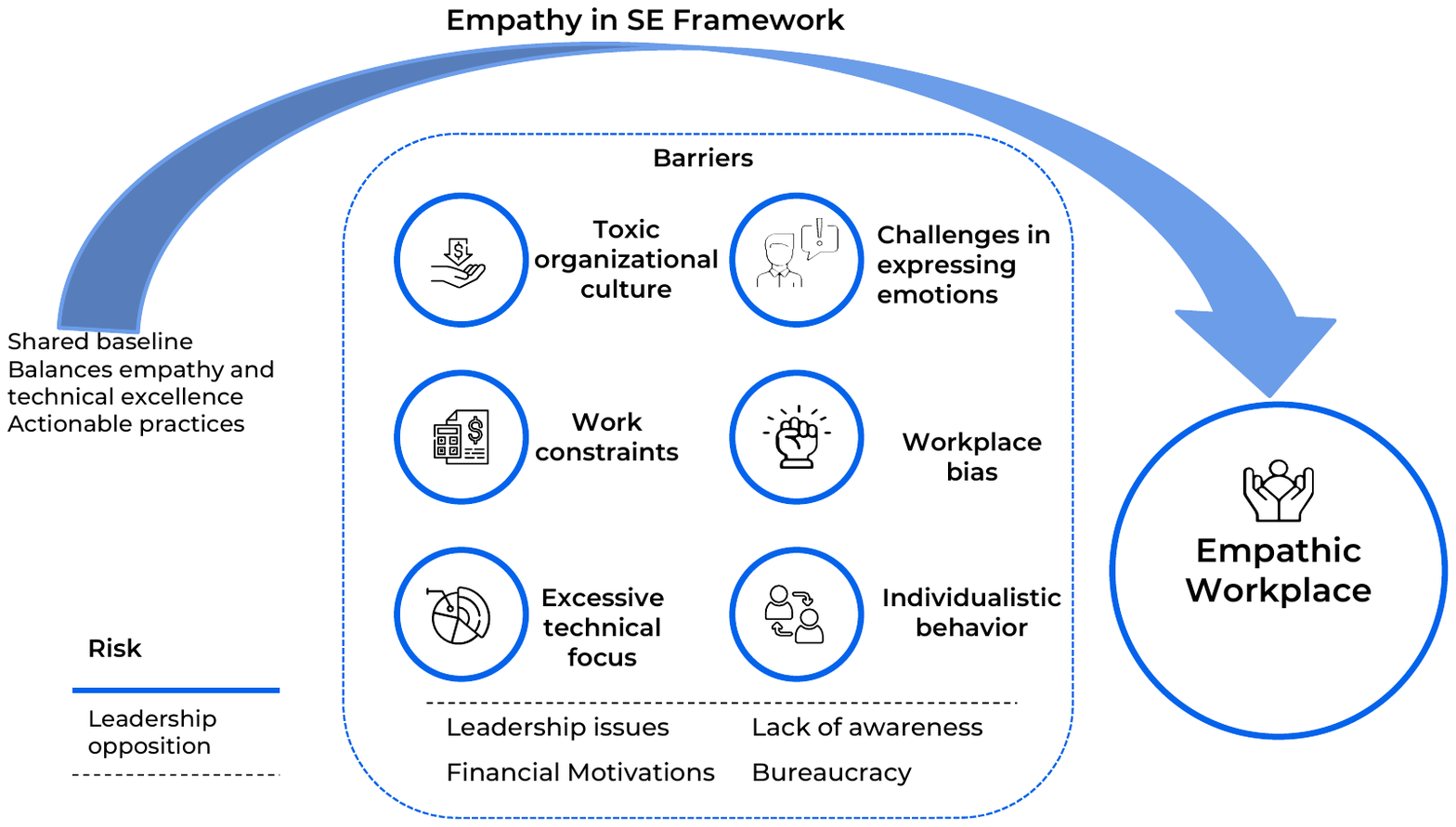}}
\caption{Empathy experts' perception of the framework's impact on the barriers to empathy in SE workplaces.}
\Description[Practitioners' perception of the framework impact on the barriers to empathy in SE workplaces.]{}
\label{fig:survey_infographic}
\end{figure}

When asked how they would address those barriers to empathy in the workplace in Q23, participants suggested multiple approaches to addressing them. Some emphasized the need for bottom-up influence, stating that CEO and senior leadership support is crucial for fostering an empathic culture and promoting organizational change. For instance, one participant stated: ``\textit{Much of this overlaps with organizational change, which an individual contributor might not have enough influence to make substantive change.}'' Another highlighted the importance of enforcing a customer focus, using structured approaches to balance technical problem-solving with user needs. The role of rewarding empathic behaviors was also noted, as workplaces that prioritize individual achievements over team success can become toxic. One participant stressed that organizations need context-specific solutions, as barriers to empathy vary depending on the workplace: ``\textit{Each one must be identified, the root causes identified and then addressed. Each organization is different}.'' Understanding cultural differences was another key factor, particularly for those working in international and multicultural teams. A few respondents believed empathy can help overcome these barriers if applied effectively. Lastly, one participant advocated for role modeling empathic behavior instead of imposing rigid rules, which can create resistance. Figure \ref{fig:survey_infographic} summarizes their perception of the framework's impact on the barriers to empathy in SE workplaces.

In Q24, 57.14\% of participants pointed out toxic organizational culture as the most challenging barrier, cited as difficult or impossible to overcome. The difficulty in expressing emotions was also seen as a significant challenge, as mentioned by 42.86\% of participants. Other barriers included excessive technical focus and work constraints (each cited once). However, 28.57\% of participants stated that none of the barriers were impossible to overcome, indicating that some practitioners believe workplace improvements are achievable despite existing challenges. 

In the answers to Q25, participants identified several reasons why barriers to empathy persist despite efforts to address them. Four respondents pointed to \textbf{leadership issues}, with some describing immature leadership that lacks the emotional intelligence needed to foster an empathic culture. Others stated that leadership actively sustains toxic workplace environments by being unaware of the negative incentives they create or benefiting from the existing culture. As one answer states: \textit{"Leaders set the tone for culture. While individual teams can find ways to carve out a haven for themselves, it is hard to make change when executives reward unempathetic behaviors."} \textbf{Financial motivations} were also a key factor, as companies often prioritize sales and performance metrics over employee well-being. \textbf{Bureaucracy} was another challenge, with participants noting that corporate structures with too many levels of management can stifle meaningful change. Additionally, one participant mentioned the 
\textbf{lack of awareness} about the importance of empathy to SE: ``\textit{Because not everyone recognizes why empathy is important. They view software engineering as a technical problem, but the truth is that technical issues rarely cause projects to fail; they are issues within the team.}'' It contributes to the problem, as some leaders and organizations view the field purely as a technical discipline, ignoring the interpersonal factors that impact project success.

The responses to Q26 and Q27 reveal mixed perceptions regarding how the practices listed in the framework can help overcome those barriers. Four participants believed the framework's practices could help overcome barriers, while three disagreed. Some participants believe the framework helps by \textbf{providing a shared baseline} that aligns understanding across teams and by \textbf{balancing technical excellence with empathy}, countering the misconception that the two are mutually exclusive. As one respondent states: ``\textit{Yes, because it sets a baseline where everyone understands what needs to be done.}'' However, two participants disagreed and expressed skepticism, arguing that \textbf{leadership can undermine efforts}, making systemic change difficult without higher accountability. Additionally, some respondents highlight the \textbf{need for further validation} of the framework and suggest making its connections to overcoming barriers more explicit in future iterations.

Lastly, the feedback on Q28 highlights both encouragement and areas for further refinement. Some participants recognize the framework's potential, emphasizing its value in demonstrating empathy’s role beyond user experience, extending to team cohesion and performance. For instance, one respondent states: ``\textit{Your framework conveys that empathy isn't only important when considering our users, but is also a critical ingredient in team cohesion and performance}.'' Others suggest future research directions, such as emphasizing measurable benefits like productivity and conducting more discussions with developers to refine the framework. Additionally, while some critique its current state, they still express support and encouragement, underscoring its promise with further development.

\textbf{Experts Takeaways. }Overall, experts view the framework as promising, well-articulated, and complete. They acknowledge its value in raising awareness about the importance of empathy in SE and in addressing the associated challenges. Furthermore, they emphasize the significance of integrating the framework into training materials and practical exercises to promote empathy-driven behaviors. Nonetheless, the experts also identified several areas for improvement, particularly the need for stronger leadership support, empirical validation, and more clearly defined practical applications.

%% file: content/5.4_limitations.tex
According to \citet{lenberg2024qualitative}, the typical internal/external/construct is better suited for statistical studies than qualitative research. We adopted the Total Quality Framework (TQF) approach, which derives from experiences in qualitative research~\cite{roller2015applied}. 
Following, we discuss the credibility (concerning data collection), analyzability (data analysis), transparency (reporting), and usefulness (results) of the study.

To ensure the \textbf{credibility} of the study, we focused on completeness and accuracy in the data collection process. The research was confined to a representative sample of articles from the DEV and Medium communities, and we acknowledge that analyzing content from other platforms could yield different insights. Consequently, we do not claim our findings to be exhaustive, as they are specific to the sample we analyzed. Besides, we discussed the planning and search process and presented our search string and article selection criteria in Appendix \ref{sec:data_collection}. We also compared our results with related papers, showing that the results are aligned with what has already been reported.   

Research bias can influence the accuracy of the \textbf{analysis}, as interpretations rely on human judgment. Two researchers selected the articles independently to reduce this risk, following the exclusion criteria. We also included a calibration step, and two independent researchers coded the data to minimize potential bias. The analysis was further discussed with senior researchers until we reached saturation. Additionally, we verified inter-coder reliability both in the pilot phase and after the coding, following the guidelines of the empirical standards \cite{ralph2020empirical}. A high level of agreement was achieved between the coders (see Appendix \ref{sec:inter-coder}). Nevertheless, we recognize that codes remain open to interpretation; therefore, we provide our data set in the experimental package for other researchers to review and validate our analysis \cite{cerqueira_dataset_2025}.

To ensure the study's \textbf{transparency}, we provide an experimental replication package, allowing others to review our data and validate our analysis. The replication package offers an audit trail of the research. We also provide the scripts used in the data collection (see Appendix \ref{sec:data_collection}), enabling replication.

Regarding the \textbf{usefulness} of our findings, our work offers valuable insights for both researchers and practitioners by advancing an under-researched topic in SE. The insights support software practitioners in overcoming empathic barriers, developing empathy skills, and fostering empathetic practices. Additionally, the research highlights the relevance of studying human factors in SE. The framework also helps disseminate the results to practitioners.

Despite the limitations to generalizability, we provide readers with enough evidence to evaluate the relevance and transferability of our findings. Additionally, to mitigate limitations from our initial study and strengthen the quality of our findings, we conducted a follow-up survey with empathy experts. This step enhanced the credibility of our work by validating and triangulating 
the results obtained from the GL analysis with firsthand data from practitioners. It also supported analyzability by providing structured responses that allowed us to examine our data. Finally, the survey also contributed to the usefulness of the research by gathering practitioners' perceptions, making our framework more grounded.  

To address potential misunderstandings (construct validity), we refined the questionnaire through expert validation, piloting, and revisions. Coding of open-ended responses introduced some subjectivity (conclusion validity), which we mitigated through a dual-coding process. Additionally, as we selected participants from DEV and Medium (external validity) at our convenience, the findings cannot be generalized. 
However, their demonstrated experience and interest in empathy in SE lend credibility to their insights.

%% file: content/6_conclusion.tex
This work uses DEV and Medium as data sources to explore empathy, an under-researched topic within SE. These platforms enabled us to access diverse practitioner perspectives, supporting a richer understanding of empathy in SE. Empathy plays a vital role in SE. However, the interest in it is still growing. In this paper, we conduct a content analysis to deepen the comprehension of this human skill. We present the following contributions. First, we conceptualize empathy from the perspective of software practitioners. Second, we describe the barriers to empathy in SE. Third, we identified empathic practices and their effects on SE. Following, we propose a conceptual framework for capturing the meaning, practices, and effects of empathy in SE and discuss how it can help overcome the barriers to empathy in SE. Moreover, we discuss how practitioners can adopt the framework in their context and discuss the benefits of such a framework.

The framework conceptualizes empathy and presents guidance for practicing empathy in SE. Although we do not claim that this perspective is the only way to define empathy, this perspective is promising based on our findings. Besides, it can help to promote a paradigm shift in SE by raising awareness of the topic. Next, we surveyed empathy experts to assess the framework. They identified addressable and persistent barriers to empathy in SE, often linked to leadership and organizational culture. Many believed these barriers could be tackled through local team practices, though systemic challenges remain. They found the framework helpful in providing shared language and promoting a balance between empathy and technical goals. Their feedback highlighted the framework’s potential while encouraging further validation and refinement.

\textit{Implications.} For practitioners, the framework can be a flexible guide to strengthen empathy in daily software tasks, from code reviews, coding, mentoring, and user-centered design. It can help practitioners develop empathy in targeted ways as it provides a visual guide for cultivating empathy in teams by identifying which empathic practices are most frequently associated with specific outcomes, offering actionable insights for improving both human and technical aspects of work.
Insights from GL and expert surveys show its potential to improve communication, code maintainability, and work culture. It serves as a learning tool for newcomers and a strategy enhancer for experienced professionals. During onboarding, it can help new hires understand team norms and foster respectful collaboration. Mentors can also use it as a reflective tool to guide mentees in developing skills critical for team success. Organizations may adopt it to foster empathy at the individual and team levels. Future technical innovations could embed empathy cues directly into SE tools to support day-to-day collaboration. For example, development environments and team dashboards could provide compassionate prompts that encourage respectful feedback during code reviews or flag potentially harsh language. AI-driven support systems might offer contextual empathy suggestions, such as highlighting stakeholder perspectives in user stories, suggesting inclusive language, or reminding developers to consider maintainability for future teammates. These features could help cultivate empathic behavior as an integrated, unobtrusive part of the SE workflow. For researchers, our work contributes to the existing body of knowledge, paving the way for new empirical studies and enhanced understanding of the topic. The research strategy we adopted can be used both for research in the same data sources and expanded to other online communities. It can also be adapted to explore other topics and concepts. Researchers can replicate or expand our approach across other online communities or related topics. The framework can be extended with new practices, scenarios, or effects. 

\textit{Future Work. } This study opens several avenues for future research to deepen our understanding of empathy in SE. One promising direction is exploring how our conceptual framework operates in agile environments, identifying opportunities to enhance collaboration, and addressing challenges in sprint planning, retrospectives, and teamwork. Another area involves examining the impact of cultural diversity on empathy within global teams, particularly how cultural norms and communication styles influence its expression and perception. Investigating empathy fatigue, especially in emotionally demanding roles like team leads or mentors, is critical to mitigating burnout and promoting well-being. 

Additionally, identifying barriers to empathy in remote and hybrid work contexts, or agile workflows, can inform strategies to address specific challenges like virtual communication, time zone differences, focus on speed, and frequent deliverables. Recognizing the effects of empathetic practices on team dynamics, job satisfaction, and project success further highlights their importance in SE. We plan to refine further the framework to incorporate the emotional regulation facet and elements of self-compassion and consider potential biases. We also recognize the potential to compare our framework to empathy-related models from design thinking, which could further support its application in human-centered software development.
Finally, developing robust metrics and evaluation methods to measure the impact of empathy in software development environments can provide actionable insights. Future work could also include statistical analysis with larger, diverse samples to validate findings. Developing tools or training based on the framework may also support adoption and evaluation. Industry case studies and controlled experiments can test the framework's real-world impact. 
These opportunities encourage interdisciplinary research and practical applications that can benefit software practitioners, teams, and organizations.

%% file: content/7_appendix_search.tex
\section{Data Collection and Selection Procedures}
\label{appendix_data_collection} 

\label{sec:data_collection}
To improve readability, this appendix presents extended methodological details regarding data collection procedures, tools used, data retrieval, and selection. We describe the data collection phase of the study, as illustrated in Figure \ref{fig:search_process}.

\subsection{Defining the search strings}  
\begin{figure}[htb]
\centerline{\includegraphics[scale=0.5, trim=0cm 5cm 0cm 7cm, clip]{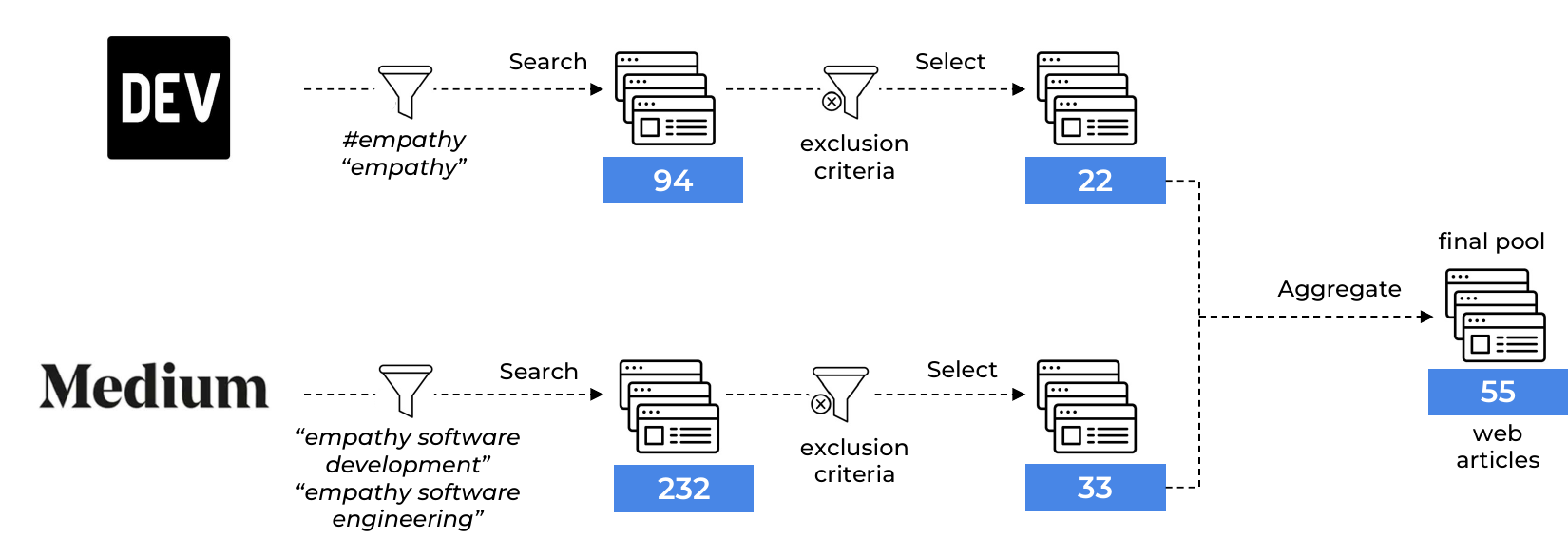}}
\caption{Collecting the web articles.}
\label{fig:search_process}
\Description[Collecting the web articles.]{Collecting the web articles.}
\end{figure}

The first step in the search process involved defining the search strings. We began by identifying terms aligned with our research questions (RQs) and then conducted a series of tests and iterative refinements to enhance the relevance of the retrieved results. For example, DEV is a community composed primarily of software practitioners; therefore, we excluded generic terms such as ``software engineering'' or ``development'' from the search string, relying solely on the keyword ``empathy.'' In contrast, when searching the Medium platform, we included the terms ``software engineering'' and ``software development'' to restrict the scope to software-related content. Across both platforms, we deliberately focused our search on the explicit term ``empathy'' to ensure the relevance and clarity of the initial dataset. This approach aimed to establish a robust foundation for understanding how software professionals perceive and discuss empathy within software engineering contexts. Table \ref{tab:search_strings} details the search strings and results. Figure \ref{fig:search_process} illustrates the search step.

\begin{table}[htb]
\caption{Search Strings.}
\begin{tabular}{lllc}
\hline
\#  & Source & Search String  & Results                    \\ \hline
1 & DEV    & \#empathy       & 34                     \\
2 & DEV    & empathy         & 60                  \\
3 & Medium & "empathy AND software AND engineering" & 183   \\
4 & Medium & "empathy AND software AND development"  & 49 \\ \hline
\end{tabular}
\label{tab:search_strings}
\end{table}

\subsection{Retrieving the data}

We executed each search string on the selected platforms and aggregated the results. Custom scripts were developed to automate the collection process and store the retrieved data in a Google Spreadsheet. All scripts used for retrieving data are available in our experimental package \cite{cerqueira_dataset_2025}.

Articles from the DEV platform were retrieved between October 10 and 31, 2022. Data collection was conducted using Forem\footnote{\url{https://docs.forem.com/api/}{https://docs.forem.com/api/}}, the official DEV API, which enables content retrieval based on user-defined tags, a common feature in DEV articles. Using the tag \#empathy (as indicated in line \#1 of Table~\ref{tab:search_strings}), we retrieved 34 articles along with their corresponding metadata. In addition to the tag-based search, we performed a content-based crawl of the DEV platform to identify articles that mentioned the term ``empathy'' (line \#2 of Table~\ref{tab:search_strings}). Due to limitations in DEV's search functionality, which returns only the top 60 results, we were able to retrieve 60 articles using this method.  

For the Medium platform, data collection was conducted between May 31 and June 1, 2023. As indicated in Table~\ref{tab:search_strings}, two search strings were employed. The search string shown in line \#3 yielded 183 articles, while the search string in line \#4 returned 49 articles. Due to Medium’s restrictions on automated web scraping, we used a browser extension\footnote{\url{https://webrobots.io/}} to manually extract the URLs and titles of the articles corresponding to each search string.

In total, we collected 94 articles from DEV and 232 from Medium, resulting in an initial dataset comprising 326 articles.

\subsection{Selecting the data}

\begin{figure}[htb]
\centerline{\includegraphics[scale=0.5, trim=0cm 3cm 0cm 3cm, clip]{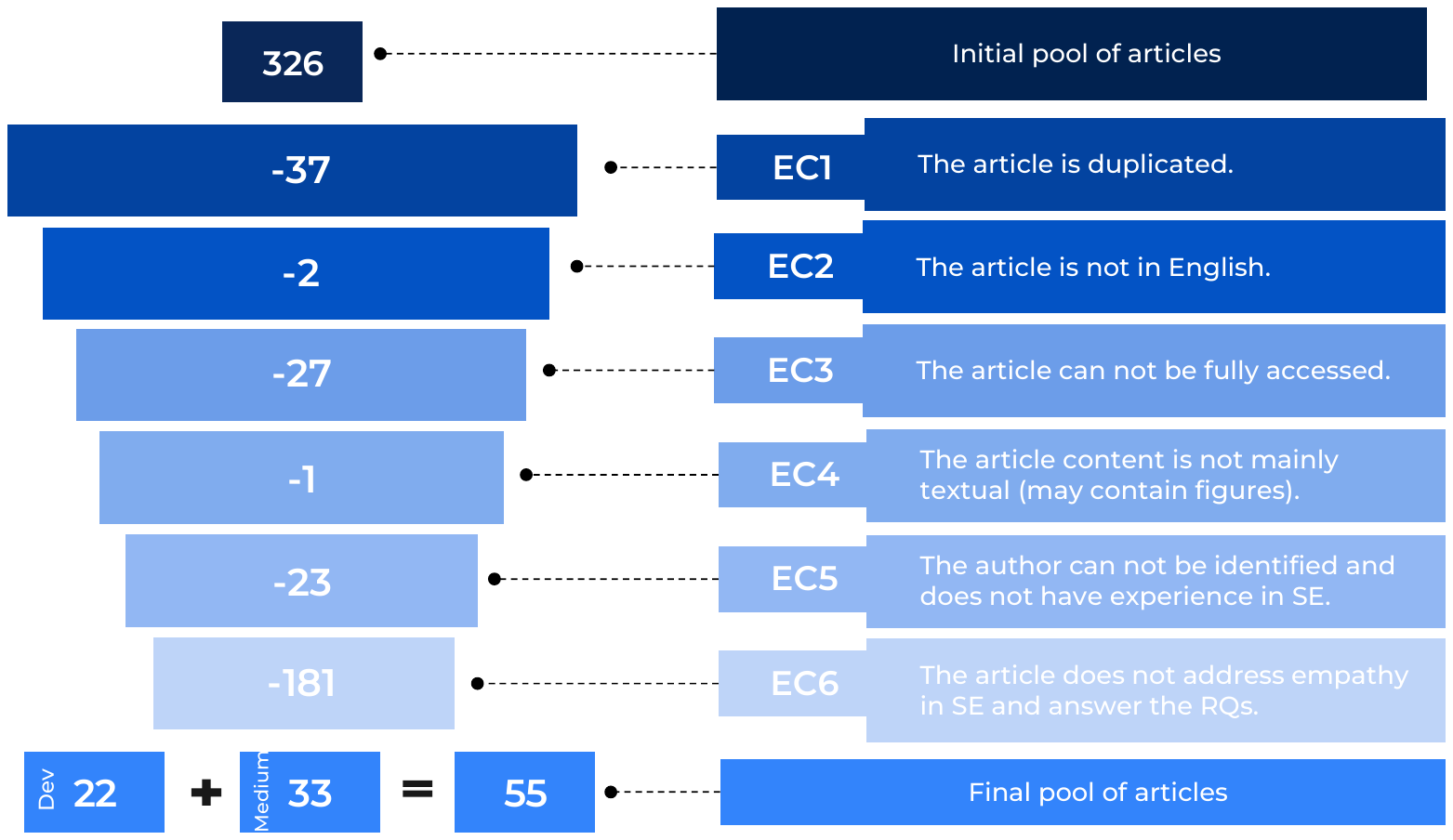}}
\caption{Applying the exclusion criteria to select the data.}
\label{fig:exclusion}
\Description[Applying the exclusion criteria to select the data.]{Applying the exclusion criteria to select the data.}
\end{figure}

We defined six exclusion criteria to filter the relevant articles among previously collected articles. 
Two independent researchers selected the texts by applying the exclusion criteria to filter the sources that did not meet them. A third researcher was consulted in case of disagreement.
Figure \ref{fig:exclusion} summarizes this process, indicating the number of sources discarded through each step of the data selection process.

We began the screening process by identifying and removing duplicate articles (EC1). Next, we verified that the articles were written in English (EC2). Articles that were not fully accessible, such as those requiring payment, were subsequently excluded (EC3). To ensure sufficient textual content for qualitative analysis, we discarded articles whose content was not predominantly textual (EC4). We then evaluated the authorship of the remaining articles, excluding those whose authors could not be identified or lacked the necessary expertise to discuss empathy within the software engineering domain (EC5). To assess author expertise, we manually reviewed authors’ GitHub and LinkedIn profiles to verify their professional or educational background related to software engineering. Articles authored by individuals without relevant experience or training in the field were excluded. This criterion was applied to ensure that the gained insights were derived from practitioners or experts with practical involvement in software engineering, thereby enhancing the reliability and contextual relevance of our findings. Finally, we evaluated whether the articles addressed empathy in software engineering and contributed to answering at least one of our research questions; those that did not meet this criterion were excluded (EC6). 

\pagebreak

%% file: content/7.1_appendix_reliability.tex
\section{Inter-coder reliability results}
\label{appendix-reliability}
In this section, we provide a broad discussion of the pilot inter-coder reliability analysis. We used the online tool Recal2~\footnote{\url{http://dfreelon.org/recal/recal2.php}} to calculate Scott’s Pi, Cohen’s Kappa, and Krippendorff’s Alpha~\cite{krippendorff2004measuring,freelon2010recal}. 

 Our pilot inter-coder reliability results show strong agreement between coders, indicating a consistent coding process (see Table \ref{tab:pilot}). Most variables exhibit 100\% agreement, with Scott’s Pi, Cohen’s Kappa, and Krippendorff’s Alpha values equal to 1. The few variables showing slightly lower agreement still indicated substantial agreement among the coders. For instance, variables 21 and 22 have 90\% agreement, with Scott’s Pi, Cohen’s Kappa, and Krippendorff’s Alpha values around 0.78–0.79. Even so, it still reflects a good level of reliability, but it indicates some minor discrepancies between coders on these variables. Variable 29 has 90\% agreement, with Scott’s Pi at 0.733, Cohen’s Kappa at 0.737, and Krippendorff’s Alpha at 0.747. Although slightly lower than the other variables, these values still indicate substantial agreement between coders.

\begin{table}[htb]
\centering
\caption{Pilot Inter-coder Reliability Results.}
\label{tab:pilot}
\small
\resizebox{0.9\columnwidth}{!}{
\begin{tabular}{cccccccc}
\bottomrule
\textbf{Variable} & \textbf{\% Agreement} & \textbf{Scott's $\pi$} & \textbf{Cohen's $\kappa$} & \textbf{Krippendorff's $\alpha$} & \textbf{N Agreements} & \textbf{N Disagreements} &  \\ \midrule
1  & 100 & 1 & 1 & 1 & 10 & 0  \\
2  & 100 & 1 & 1 & 1 & 10 & 0  \\
3  & 100 & 1 & 1 & 1 & 10 & 0  \\
4  & 100 & 1 & 1 & 1 & 10 & 0 \\
5  & 100 & 1 & 1 & 1 & 10 & 0  \\
6  & 100 & 1 & 1 & 1 & 10 & 0  \\
7  & 100 & 1 & 1 & 1 & 10 & 0  \\
8  & 100 & 1 & 1 & 1 & 10 & 0  \\
9  & 100 & 1 & 1 & 1 & 10 & 0 \\
10 & 100 & 1 & 1 & 1 & 10 & 0  \\
11 & 100 & 1 & 1 & 1 & 10 & 0  \\
12 & 100 & 1 & 1 & 1 & 10 & 0  \\
14 & 100 & 1 & 1 & 1 & 10 & 0  \\
16 & 100 & 1 & 1 & 1 & 10 & 0  \\
20 & 100 & 1 & 1 & 1 & 10 & 0  \\
21 & 90  & 0.7802 & 0.7826 & 0.7912 & 9  & 1  \\
22 & 90  & 0.7802 & 0.7826 & 0.7912 & 9  & 1  \\
23 & 100 & 1 & 1 & 1 & 10 & 0  \\
24 & 100 & 1 & 1 & 1 & 10 & 0  \\
27 & 100 & 1 & 1 & 1 & 10 & 0  \\
28 & 100 & 1 & 1 & 1 & 10 & 0  \\
29 & 90  & 0.733 & 0.737 & 0.747 & 9  & 1  \\
30 & 100 & 1 & 1 & 1 & 10 & 0  \\
31 & 100 & 1 & 1 & 1 & 10 & 0  \\
32 & 100 & 1 & 1 & 1 & 10 & 0  \\
33 & 100 & 1 & 1 & 1 & 10 & 0  \\
36 & 100 & 1 & 1 & 1 & 10 & 0  \\
40 & 100 & 1 & 1 & 1 & 10 & 0  \\
42 & 100 & 1 & 1 & 1 & 10 & 0  \\
44 & 100 & 1 & 1 & 1 & 10 & 0  \\
51 & 100 & 1 & 1 & 1 & 10 & 0  \\
52 & 100 & 1 & 1 & 1 & 10 & 0 \\
53 & 100 & 1 & 1 & 1 & 10 & 0 \\
54 & 100 & 1 & 1 & 1 & 10 & 0  \\
55 & 100 & 1 & 1 & 1 & 10 & 0  \\
\midrule
\multicolumn{4}{l}{\textbf{\small{Caption:}}} \\
\multicolumn{8}{l}{\small{\textbf{Variable:} Lists the assessed variables by number.}} \\
\multicolumn{8}{l}{\small{\textbf{\%Agreement:} Shows the percentage of instances where coders agreed.}}\\
\multicolumn{8}{l}{\small{\textbf{Scott's $\pi$:} Measures agreement while accounting for a chance.}}\\
\multicolumn{8}{l}{\small{\textbf{Cohen's $\kappa$:} Similar to Scott's $\pi$, evaluating agreement beyond chance.}}\\
\multicolumn{8}{l}{\small{\textbf{Krippendorff’s $\alpha$:} Assesses reliability across different measurement levels.}}\\
\multicolumn{8}{l}{\small{\textbf{N Agreements:} Number of instances where coders agreed.}}\\
\multicolumn{8}{l}{\small{\textbf{N Disagreements:} Number of instances where coders disagreed.}}\\

\bottomrule
\end{tabular}
}
\end{table}


The final inter-coder reliability results presented in Table \ref{tab:reliability-analysis} demonstrate high levels of agreement in multiple variables. It is generally above 90\%. The strong reliability reflected by Cohen's Kappa, Scott's Pi, and Krippendorff's Alpha further confirms the high consistency of the coding decisions. Most variables show values above 0.80 for all three metrics, indicating substantial to almost perfect agreement. The metrics are slightly lower for variables 15, 16, and 20 than for other variables. However, values around 0.73 to 0.79 still indicate substantial agreement between the coders. The minimal disagreements in most variables highlight the process consistency.

\begin{table}[htb]
\centering
\caption{Final Inter-coder Reliability Results.}
\label{tab:reliability-analysis}
\resizebox{0.9\columnwidth}{!}{
\begin{tabular}{cccp{2cm}p{3cm}ccc}
\hline
\textbf{Variable} & \textbf{\% Agreement} & \textbf{Scott's $\pi$} & \textbf{Cohen's $\kappa$} & \textbf{Krippendorff's $\alpha$} & \textbf{N Agreements} & \textbf{N Disagreements} &  \\ \midrule
 1  & 94.55  & 0.89  & 0.89  & 0.89  & 52  & 3    \\    
2  & 94.55  & 0.86  & 0.86  & 0.86  & 52  & 3    \\    
3  & 92.73  & 0.80  & 0.80  & 0.80  & 51  & 4    \\    
4  & 94.55  & 0.84  & 0.84  & 0.84  & 52  & 3    \\    
5  & 98.18  & 0.94  & 0.94  & 0.94  & 54  & 1    \\    
6  & 96.36  & 0.93  & 0.93  & 0.93  & 53  & 2    \\    
7  & 98.18  & 0.95  & 0.95  & 0.95  & 54  & 1    \\    
8  & 98.18  & 0.95  & 0.95  & 0.95  & 54  & 1    \\    
9  & 98.18  & 0.94  & 0.94  & 0.94  & 54  & 1    \\    
10 & 100.00 & 1.00  & 1.00  & 1.00  & 55  & 0    \\    
11 & 96.36  & 0.81  & 0.81  & 0.81  & 53  & 2    \\    
12 & 98.18  & 0.88  & 0.88  & 0.88  & 54  & 1    \\    
13 & 100.00 & 1.00  & 1.00  & 1.00  & 55  & 0    \\    
14 & 96.36  & 0.87  & 0.87  & 0.87  & 53  & 2    \\    
15 & 98.18  & 0.79  & 0.79  & 0.79  & 54  & 1    \\    
16 & 96.36  & 0.73  & 0.73  & 0.73  & 53  & 2    \\    
17 & 98.18  & 0.88  & 0.88  & 0.88  & 54  & 1    \\    
18 & 100.00 & 1.00  & 1.00  & 1.00  & 55  & 0    \\    
19 & 100.00 & 1.00  & 1.00  & 1.00  & 55  & 0    \\    
20 & 98.18  & 0.79  & 0.79  & 0.79  & 54  & 1    \\    
21 & 92.73  & 0.85  & 0.85  & 0.85  & 51  & 4    \\    
22 & 94.55  & 0.84  & 0.84  & 0.84  & 52  & 3    \\    
23 & 98.18  & 0.94  & 0.94  & 0.94  & 54  & 1    \\    
24 & 94.55  & 0.84  & 0.84  & 0.84  & 52  & 3    \\    
25 & 98.18  & 0.91  & 0.91  & 0.91  & 54  & 1    \\    
26 & 96.36  & 0.85  & 0.85  & 0.85  & 53  & 2    \\    
27 & 100.00 & 1.00  & 1.00  & 1.00  & 55  & 0    \\    
28 & 100.00 & 1.00  & 1.00  & 1.00  & 55  & 0    \\    
29 & 98.18  & 0.88  & 0.88  & 0.88  & 54  & 1    \\    
30 & 98.18  & 0.88  & 0.88  & 0.88  & 54  & 1    \\    
31 & 100.00 & 1.00  & 1.00  & 1.00  & 55  & 0    \\    
32 & 100.00 & 1.00  & 1.00  & 1.00  & 55  & 0    \\    
33 & 100.00 & 1.00  & 1.00  & 1.00  & 55  & 0    \\    
34 & 100.00 & 1.00  & 1.00  & 1.00  & 55  & 0    \\    
35 & 100.00 & 1.00  & 1.00  & 1.00  & 55  & 0    \\    
36 & 100.00 & 1.00  & 1.00  & 1.00  & 55  & 0    \\    
37 & 100.00 & 1.00  & 1.00  & 1.00  & 55  & 0    \\    
38 & 100.00 & 1.00  & 1.00  & 1.00  & 55  & 0    \\    
39 & 100.00 & 1.00  & 1.00  & 1.00  & 55  & 0    \\    
40 & 100.00 & 1.00  & 1.00  & 1.00  & 55  & 0    \\    
41 & 100.00 & 1.00  & 1.00  & 1.00  & 55  & 0    \\    
42 & 100.00 & 1.00  & 1.00  & 1.00  & 55  & 0    \\    
43 & 100.00 & 1.00  & 1.00  & 1.00  & 55  & 0    \\    
44 & 100.00 & 1.00  & 1.00  & 1.00  & 55  & 0    \\    
45 & 98.18  & 0.95  & 0.95  & 0.95  & 54  & 1    \\    
46 & 96.36  & 0.90  & 0.90  & 0.90  & 53  & 2    \\    
47 & 98.18  & 0.91  & 0.91  & 0.91  & 54  & 1    \\    
48 & 98.18  & 0.94  & 0.94  & 0.94  & 54  & 1    \\    
49 & 98.18  & 0.93  & 0.93  & 0.93  & 54  & 1    \\    
50 & 98.18  & 0.93  & 0.93  & 0.93  & 54  & 1    \\    
51 & 96.36  & 0.87  & 0.87  & 0.87  & 53  & 2    \\    
52 & 98.18  & 0.93  & 0.93  & 0.93  & 54  & 1    \\    
53 & 98.18  & 0.94  & 0.94  & 0.94  & 54  & 1    \\    
54 & 100.00 & 1.00  & 1.00  & 1.00  & 55  & 0    \\    
55 & 100.00 & 1.00  & 1.00  & 1.00  & 55  & 0    \\    
56 & 98.18  & 0.94  & 0.94  & 0.94  & 54  & 1    \\    
\bottomrule
\end{tabular}
}

\end{table}

\pagebreak

%% file: content/7.2_appendix_extraction.tex
\clearpage

\section{Data extraction and synthesis}
\label{appendix-synthesis}

In this section, we detail the data extraction and synthesis process, using an illustrative example.

\subsection{Data extraction} 

Following the pilot analysis, we proceeded with the full data analysis. The two coders independently extracted relevant quotes from each article to address the research questions. They identified and selected meaningful text passages based on their own interpretations, without the use of a pre-established coding scheme. This open coding approach was adopted due to the lack of prior analyses on the selected articles and the exploratory nature of our study. It enabled a flexible yet systematic data acquisition process, ensuring that any information potentially relevant to the research questions was captured while maintaining the interpretive depth of the qualitative analysis.

Figure~\ref{fig:coding} presents an example of data extraction from a selected article. In this example, practitioner P30\footnote{Each practitioner author is referenced using a unique identifier, such as P1, P2, ..., Pn.} reflects on the economics of empathy within the context of software engineering\footnote{\href{https://medium.com/samsung-internet-dev/the-economics-of-empathy-b4d49362b7b7}{The Economics of Empathy}}. The highlighted excerpts shown are illustrative quotes from our coded dataset, each marked and color-coded according to the research question it addresses.

To elicit RQ1, the meaning of empathy, we quoted: \textit{``understand the feelings and experiences of others''}. To elicit RQ2, we quoted the following excerpt to extract the barriers to empathy: \textit{``When we’re developing software, it is easy to get carried away with the tech...We sometimes forget that we’re building things that a whole spectrum of people will use.''} To elicit RQ3, we quoted the empathetic practices mentioned by P30: \textit{``Understanding their needs''} and \textit{``running user testing''}. To elicit RQ4, we extracted the effects of practicing empathy. For instance, \textit{``building software that they enjoy using''}.

The extraction was conducted through peer review. The text quotes, annotations, and comments were registered on Google Sheets spreadsheets, ensuring traceability links between the extracted data and the primary sources. After finishing each iteration of the extraction, we systematically compared the results, resolving disagreements in consensus meetings.

\subsection{Synthesis}

Figure \ref{fig:coding} shows how we organized the quotes into descriptive labels that answered the RQs. 
Two researchers revised and unified the codes during each analysis cycle until they did not identify new codes. Disagreements were resolved in consensus meetings. A third researcher revised the codes and relationships. At the end of this process, the codebook was built. 

\subsubsection{Open coding}

\begin{figure}[t]
\centerline{\includegraphics[width=\linewidth, trim=0cm 3cm 5cm 3cm, clip]{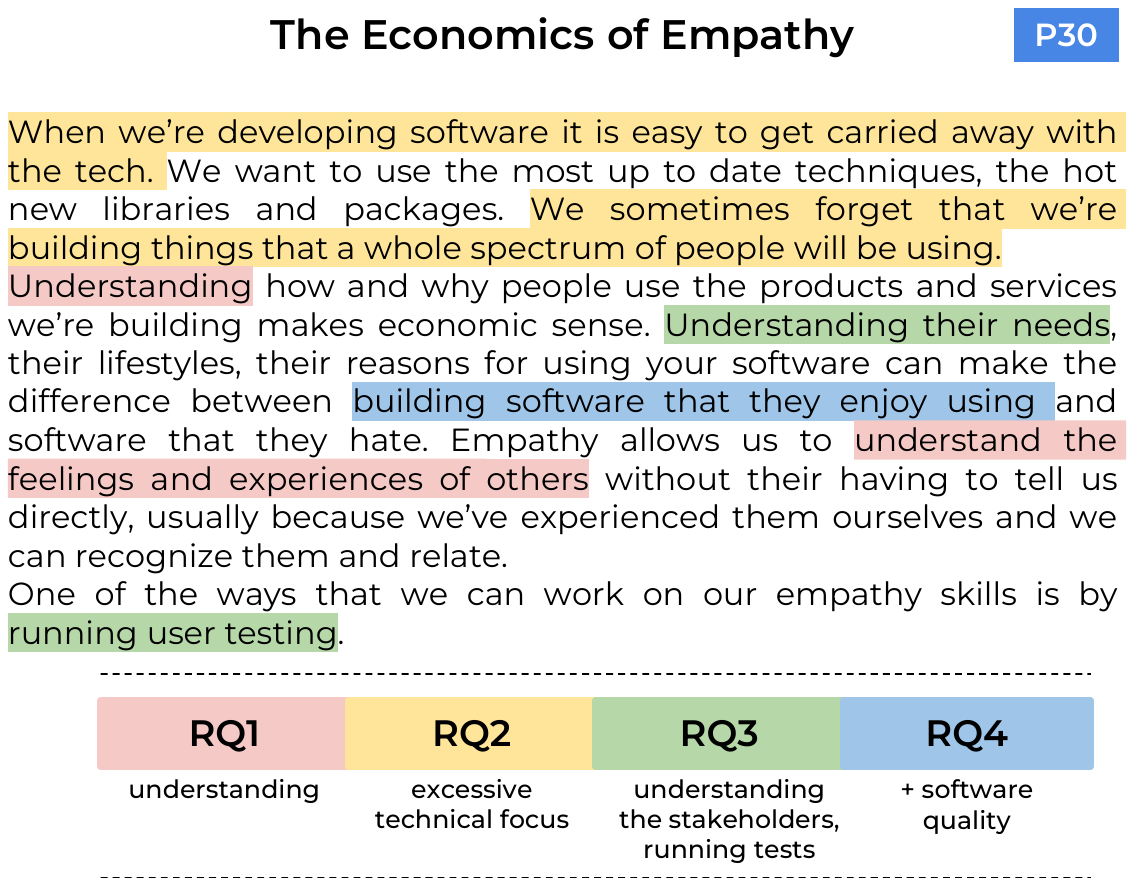}}
\caption{Answering our RQs with an article from Medium. In this excerpt, we show how we extract answers from RQ1-RQ4.}
\label{fig:coding}
\Description[Answering our RQs with an article from Medium. In this excerpt, we show how we extract answers from RQ1-RQ4]{Answering our RQs with an article from Medium. In this excerpt, we show how we extract answers from RQ1-RQ4}
\end{figure}

We applied an iterative, open-coding process to synthesize the previously extracted information. In the example of Figure \ref{fig:coding}, we coded the meaning of empathy as: \textit{understanding}. Some practitioners cited more than one meaning of empathy, which we coded accordingly. At the final stage of this process, we have a list of codes that represent the meaning of empathy and its number of occurrences, supporting us in answering RQ1.  

In answering RQ1, we compared our codes with previous studies \cite{gerdes2009social,hess2016voices,walther2017model, decety2021emergence}. For example, we initially coded the statement \textit{``empathy is simply the ability to put yourself in the shoes of another human being''} as \textit{walking in others’ shoes}. After comparing our results with prior work, we refined this to \textit{embodiment}. We then organized the meanings of empathy into broader themes aligned with the facets of empathy identified by~\cite{decety2021emergence}. For instance, we grouped the codes \textit{understanding}, \textit{perspective taking}, and \textit{embodiment} under the high-level theme of \textit{cognitive empathy}.

For RQ2, we verified if the practitioners mentioned barriers to practicing empathy in SE. In Figure~\ref{fig:coding}, P30 discussed how we tend to focus on technology in software development and forget that we are building software for people. Thus, we coded this quote as \textit{ excessive focus on technical aspects}.

To answer RQ3, we looked to see if the practitioners mentioned empathetic practices. We coded the following practices described by P30 in Figure~\ref{fig:coding} \textit{understanding the stakeholders} and \textit{running tests}.

Besides, we also identified what these practices could affect and coded \textit{software quality} to answer RQ4. 

\subsubsection{Reviewing the codes}
\label{reviewing}

As part of the iterative open coding process, we reviewed the code attributed to some practices to better reflect their meaning. 

This step was used, for instance, when we revisited the DEV data analyzed in our previous studies \cite{cerqueira2023sbes,cerqueira2024empathy}. For example, we updated the following practices: \textit{testing} as \textit{running tests}, \textit{taking perspective} as \textit{considering different perspectives}, \textit{understanding others} as \textit{understanding the stakeholders}, and \textit{attentive listening} as \textit{actively listening}. 

We also consolidated certain practices that were initially coded separately. For example, the following quote was originally coded as \textit{anticipating questions} and \textit{reviewing the code}:

\begin{quote}
\textit{“Before you send any code review, take the time to scan through your code for clear issues. Avoiding the need for someone else to point out obvious issues in your code is one of the key ways you can be respectful of their time.
2. Anticipate your coworkers’ questions.”} [P16]
\end{quote}

Upon further review, we determined that both actions reflect a common intent and context, preparing one's code thoughtfully before submission. As a result, we merged the two codes and retained only \textit{reviewing the code} as the representative practice.

\subsubsection{Final inter-coder reliability measuring}
\label{sec:inter-coder}

After completing the full coding process, we recalculated inter-coder reliability to assess the consistency of our analyses (see Appendix~\ref{appendix-reliability}). Consistent with the pilot phase, the metrics reported in Table~\ref{tab:reliability-analysis} demonstrate high levels of agreement across most coding categories. They suggest a robust alignment and underscore the coders' strong shared understanding of the coding scheme. The agreement in multiple variables, generally above 90\%, and the strong reliability further confirm the high consistency of the coding decisions. 

Comparing the pilot (see section \ref{sec:pilot}) and final inter-coder reliability results, we observe a consistent and high level of agreement across both steps. But, while the reliability results of the pilot analysis showed minor discrepancies in a few variables, the final analysis demonstrates improvement and stability in the process. This improvement suggests that initial ambiguities were effectively resolved, thereby reinforcing the rigor of the coding process and bolstering confidence in the accuracy and consistency of our qualitative findings.  

\subsubsection{Relating the codes} \label{relating_codes}

Next, we investigated the relationships between coded practices and their corresponding effects. We synthesize the relationships between the practices and effects and classify the effects according to types. We categorize these effects into positive effects and reductive effects. \textbf{Positive effects} are those outcomes that actively enhance or improve certain desirable aspects of the work environment or the behavior of the team members. These effects create a more productive, motivated, and cohesive team. Conversely, \textbf{reductive effects} refer to the outcomes of a practice that reduce or mitigate negative aspects or stressors within the work environment. 

In the example of Figure \ref{fig:coding}, \textit{running tests} can positively affect the \textit{software quality} and \textit{understanding the stakeholders} has a positive effect on the \textit{work environment}. 

By categorizing effects in this manner, we create a clearer framework for understanding how empathetic practices influence both the enhancement of positive conditions and the reduction of potential impediments in SE contexts. As a result of these relationships, we built a conceptual framework of empathy in SE presented in Section \ref{sec:framework}. 

%% file: content/7.3_appendix_results.tex
\clearpage
\section{Extracted Codes}
\label{sec:app-codes}

This appendix presents the extracted codes in the content analysis of Grey Literature.

\begin{table}[htb]
\centering
\caption{The meaning of empathy according to software practitioners.}
\resizebox{\columnwidth}{!}{
\begin{tabular}{llp{4cm}cp{4cm}}
\toprule
\textbf{Facet} & \textbf{Theme} & \textbf{Description} & \textbf{Frequency} & \textbf{Practitioners} \\ \midrule
\multirow{10}{*}{Cognitive} & \multirow{4}{*}{Understanding} & The capacity to understand another person’s thoughts, needs, feelings, problems, and experience & 29 & P4, P5, P6, P8, P9, P10, P11, P12, P14, P15, P26, P27, P29, P30, P31, P32, P33, P34, P37, P38, P40, P41, P42, P45, P46, P49, P50, P51, P52
 \\ \cmidrule{2-5}
 & \multirow{3}{*}{Perspective taking} & Thinking, taking into account, imagining, or seeing another’s perspective & 18 & P1, P2, P13, P15, P16, P18, P19, P20, P31, P32, P36, P39, P41, P47, P52, P53, P54, P55 \\ \cmidrule{2-5}
 & \multirow{3}{*}{Embodiment} & Putting oneself in someone else’s shoes or in another’s place & 11 & P7, P8, P10, P14, P15, P16, P40, P41, P46, P49, P54 \\ \hline
\multirow{2}{*}{Compassionate} & \multirow{2}{*}{Compassion} & A kind response to someone’s challenges, often accompanied by a desire to support or help them & 14 & P3, P6, P9, P10, P12, P17, P20, P21, P22, P34, P43, P44, P48, P54 \\ \hline
\multirow{3}{*}{Emotional} & \multirow{3}{*}{Emotional sharing} & When one can feel another person’s emotions or share the emotional state of others & 13 & P10, P11, P14, P16, P25, P27, P32, P34, P38, P40, P47, P50, P54 \\ 
\bottomrule
\end{tabular}
}
\label{tab:facets}
\end{table}

\begin{table}[htb]
\centering
\caption{Barriers to empathy according to software practitioners}
\resizebox{\columnwidth}{!}{
\begin{tabular}{p{4.5cm}p{4.2cm}cp{3.5cm}}
\toprule
\textbf{Barrier} & \textbf{Description} & \textbf{Frequency} & \textbf{Mentions} \\
\midrule
\multirow{2}{*}{Toxic organizational culture} & Negative cultural aspects that affect empathy & 15 & P3, P10, P14, P15, P16, P17, P18, P23, P24, P26, P27, P34, P39, P40, P54 \\

\hline
\multirow{2}{*}{Individualistic behavior} & Lack of practitioners' effort to get the perspective of others. & 12 &  P1, P7, P11, P13, P14, P16, P29, P31, P38, P49, P54, P55\\

\hline
\multirow{2}{*}{Workplace bias} & Stereotypes and negative perceptions within the tech industry. & 11 &  P4, P5, P7, P10, P16, P18, P23, P24, P25, P28, P29\\

\hline
\multirow{2}{*}{Excessive technical focus} & Prioritizing technology over empathy.  & 10 & P1, P2, P3, P4, P12, P30, P31, P33, P41, P51\\

\hline
\multirow{3}{*}{Challenges in expressing emotions} & The challenge of understanding tone and intention in communication. & 6 & P5, P6, P14, P16, P19, P29 \\

\hline
\multirow{2}{*}{Work constraints} & The pressures developers face can hinder empathetic coding. & 2 & P5, P55 \\

\bottomrule
\end{tabular}
}
\label{tab:barriers}
\end{table}

\begin{table}[htb]
\centering
\caption{Summary of Empathetic Practices}
\resizebox{\columnwidth}{!}{
\begin{tabular}{lp{5cm}cp{3.5cm}}
\toprule
\textbf{Practice} & \textbf{Description} & \textbf{Occurrences} & \textbf{Mentions} \\ 
\midrule
Understanding the stakeholders & Understanding what stakeholders need, think, or feel & 48 & P3, P5, P9, P10, P13, P16, P19, P20, P23, P25, P26, P27, P28, P30, P31, P32, P33, P34, P37, P40, P41, P46, P49, P53, P54 \\ \hline
Considering different perspectives & Being able to take the perspective of the stakeholders & 36 & P1, P5, P7, P18, P23, P24, P28, P29, P31, P39, P41, P42, P44, P49, P50, P51, P52, P53, P55 \\ \hline
Adopting good programming practices & Thinking about future code maintainers, following best programming practices. & 28 & P2, P7, P10, P16, P19, P20, P21, P22, P26, P35, P36, P43, P50, P52, P55 \\ \hline
Being compassionate & Being polite, respectful, showing compassion, and kindness to other stakeholders.  & 24 & P4, P6, P10, P14, P17, P20, P26, P34, P36, P45, P50, P51, P52 \\ \hline
Running tests & Thinking about the users and future maintainers when testing the software and trying to eliminate bugs and errors & 15 & P2, P10, P12, P25, P30, P35, P43 \\ \hline
Actively listening & Actively listening to the needs of the stakeholders and asking meaningful questions & 15 & P8, P15, P42, P44, P46, P51, P52 \\ \hline
Being mindful & Being mindful of interactions with any of the stakeholders & 15 & P4, P11, P16, P28, P32 \\ \hline
Accepting and Giving feedback & Being open to receiving feedback and providing positive, specific, constructive, and actionable feedback to the stakeholders & 12 & P16, P26, P49, P51 \\ \hline
Being open & Communicating honestly and sharing knowledge with team members & 12 & P6, P8, P9, P14, P15, P33, P34, P44, P46 \\ \hline
Being patient & Staying calm under pressure, slowing down, offering empathy, and proceeding calmly & 12 & P15, P48, P52 \\ \hline
Taking care & Caring for stakeholders & 11 & P8, P15, P17, P27, P44, P52 \\ \hline
Admitting failures & Admitting wrong, accepting mistakes, and creating a continuous learning culture where failures are learning opportunities & 9 & P10, P15, P34, P44, P48, P53 \\ \hline
Embodying the stakeholders & Getting in the shoes of the stakeholders & 7 & P30, P38, P42, P47, P51, P53 \\ \hline
Reviewing the code & Reviewing the code before submitting it & 6 & P16, P26 \\ \hline
Being supportive & Offering support to other colleagues and team members & 4 & P1, P24, P50 \\ \hline
\end{tabular}
}
\label{tab:practices}
\end{table}

\begin{table}[htb]
\centering
\caption{Effects of practicing empathy mentioned by software practitioners}
\label{tab:empathy_effects}
\resizebox{0.9\columnwidth}{!}{
\begin{tabular}{lp{5cm}cp{5cm}}
\toprule
\textbf{Effect} & \textbf{Description} & \textbf{Frequency} & \textbf{Mentions} \\ \midrule
Software quality & Improving the code quality and creating better products & 69 & P1, P2, [P3], P4, P5, P6, P7, P8, P9, P10, P11, P12, P13, P14, P15, P16, P17, P18, P19, P20, P21, P22, P23, P24, P25, P26, P27, P28, P29, P30, P31, P32, P33, P34, P35, P36, P37, P38, P39, P40, P41, P42, P43, P44, P45, P46, P47, P48, P49, P50, P51, P52, P53, P54, P55 \\ \hline
Communication & Improving stakeholders communication & 23 & P5, P6, P12, P16, P22, P23, P25, P26, P30, P32, P41, P42, P48, P51, P52 \\ \hline
Trust & Establishing trust among team members through empathy & 23 & P1, P4, P8, P15, P16, P17, P26, P34, P44, P52 \\ \hline
Collaboration & Empathy fostering better collaboration between the stakeholders & 19 & P4, P14, P22, P26, P29, P30, P32, P34, P43, P49, P51, P52 \\ \hline
Work culture & Enhancing organizational culture with empathetic practices & 16 & P15, P17, P24, P25, P34, P35, P43, P45, P49, P50 \\ \hline
Productivity & Increasing the productivity in software development & 15 & P15, P17, P20, P22, P26, P33, P34, P50, P52, P54 \\ \hline
Work environment & Creating a supportive and empathetic workplace & 13 & P4, P14, P16, P17, P26, P29, P30, P48, P51 \\ \hline
Professional growth & Empathy facilitating career development & 8 & P1, P16, P24, P27, P40, P48 \\ \hline
Blame & Reducing blame between stakeholders & 7 & P6, P7, P13, P15, P18, P34 \\ \hline
Human connections & Building strong human connections within the team & 6 & P6, P26, P41, P52 \\ \hline
Employee retention & Retaining employees by promoting empathy in the workplace & 5 & P4, P25, P48, P50 \\ \hline
Respect & Encouraging mutual respect among peers & 5 & P9, P16, P26 \\ \hline
Team & Empathy leading to better relation between team members and results & 5 & P4, P36, P43, P52 \\ \hline
Feedback & Improving the quality of feedback & 4 & P15, P16 \\ \hline
Inclusion & Promoting inclusion & 4 & P19, P25, P28 \\ \hline
Leadership & Improving leadership & 4 & P4, P17, P30, P32 \\ \hline
Self-development & Personal growth encouraged by empathy & 4 & P24, P32, P40, P42 \\ \hline
Stress & Helping reducing stress in the workplace & 4 & P15, P48, P52 \\ \hline
Happiness & Boosting practitioners happiness & 3 & P24, P25, P52 \\ \hline
Motivation & Driving employee motivation & 3 & P4, P9, P14 \\ \hline
Anxiety & Reducing anxiety in work settings & 2 & P48 \\ \hline
Compassion & Building compassion between stakeholders & 2 & P10, P12 \\ \hline
Confidence & Boosting confidence in teams & 2 & P35 \\ \hline
Developers' onboarding & Empathy aiding in the onboarding of new developers & 2 & P2, P20 \\ \hline
Pride & Building pride in one's work & 2 & P12, P14 \\ \hline
Problem solving & Improving problem-solving abilities & 2 & P24, P41 \\ \hline
Psychological safety & Fostering psychological safety through empathy & 2 & P28, P34 \\ \hline
Frustration & Empathy reducing frustration in teams & 1 & P26 \\ \hline
Morale & Boosting the team morale  & 1 & P15 \\ 
\hline
Burnout & Preventing workers burnout & 1 & P15 \\ 
\bottomrule
\end{tabular}
}
\end{table}

%% file: content/7.4_appendix_survey.tex
\clearpage
\section{Survey with empathy experts to assess the conceptual framework}
\label{appendix-survey}

This appendix presents the full questionnaire used in the survey with empathy experts to assess the conceptual framework developed in this study. The instrument was designed to gather expert feedback on the framework’s clarity, relevance, completeness, and practical applicability in software engineering contexts.

\textbf{Section 1} contains three questions to collect participants' demographic data (see Table \ref{tab:demographics}). This section gathers background information about the participants to contextualize their responses and support the analysis.

\begin{table}[htb]
\centering
\small
\renewcommand{\arraystretch}{1.3} 
\caption{Demographic questions}
\label{tab:demographics}
\begin{tabular*}{\textwidth}{p{1cm}p{7cm}p{5cm}p{3cm}}
\hline
\textbf{No} & \textbf{Question (Q) Description} & \textbf{Options} & \textbf{Type} \\ 
\hline
Q1 & In which country do you currently work or last worked? & - & Open \\ 
\hline
Q2 & Which of the following best describes your role in your current or most recent job? 
& \begin{tabular}[t]{@{}l@{}}
Software Developer/Engineer \\ 
Quality Assurance/Tester \\ 
Project Manager \\ 
Senior Manager/Executive \\ 
Business Analyst \\ 
UX/UI Designer \\ 
DevOps Engineer \\ 
Product Manager \\ 
Other 
\end{tabular} & Closed \\ 
\hline
Q3 & How many years of professional experience do you have in your field? 
& \begin{tabular}[t]{@{}l@{}}
Less than 1 year \\ 
1–5 years \\ 
5–10 years \\ 
10–15 years \\ 
More than 15 years 
\end{tabular} & Closed \\ 
\hline
\end{tabular*}
\end{table}

\pagebreak
\textbf{Section 2} begins with one closed-ended question aimed at understanding participants' self-assessed expertise level in empathy within SE (see Table \ref{tab:section2-3}). This section sets the context for their responses and helps gauge how familiarity with the concept may influence their insights. Following this, we encourage the participants to consult the conceptual framework before answering the questions, ensuring they are familiar with its content and can provide well-informed feedback. We only invited the authors of the texts collected to build the framework. We then instructed
the practitioners to watch a 6-minute explanatory video summarizing the framework. We decided to use an explanatory video instead of text to facilitate the practitioners’
participation. Also, we provided the video transcript and a PDF file containing the framework for download. Then, 
the participants answered Q5-Q10, in \textbf{Section 3}, in which we asked if anything about their perception of empathy in SE changed what they learned and if they disagreed with any concept or relationship proposed in the framework.

\begin{table*}[htb]
\centering
\small
\renewcommand{\arraystretch}{1.3} 
\caption{Questions assessing expertise and the framework influence.}
\label{tab:section2-3}

\begin{tabular*}{\textwidth}{p{1cm}p{8cm}p{4cm}p{2cm}}
\hline
\textbf{No} & \textbf{Question (Q) Description} & \textbf{Options} & \textbf{Type} \\ 
\hline
\multicolumn{4}{l}{\textbf{Section 2. Level of expertise about empathy in SE}} \\ 
\hline
Q4 & Please indicate your level of expertise about empathy in software engineering by selecting the option that best describes your knowledge level: 
& \begin{tabular}[t]{@{}l@{}}
Novice  \\ 
Intermediate  \\ 
Advanced  \\ 
Expert
\end{tabular} 
& Closed \\ 
\hline
\multicolumn{4}{l}{\textbf{Section 3. Assess the framework influence on the perception of empathy in SE}} \\ 
\hline
Q5 & After exploring the framework, what changed in your perception of empathy in SE? For example, are there concepts or ideas you did not know about before, now think about differently, or now understand more completely? 
& - & Open \\ 
\hline
Q6 & What have you learned about empathy in SE from our framework? & - & Open \\ 
\hline
Q7 & Do you disagree with any element presented in our framework? 
& \begin{tabular}[t]{@{}l@{}}
Yes \\ 
No 
\end{tabular} 
& Closed \\ 
\hline
Q8 & If yes, how? If not, why? & - & Open \\ 
\hline
Q9 & Are there any concepts or definitions within the framework that you find confusing or misleading? 
& \begin{tabular}[t]{@{}l@{}}
Yes \\ 
No 
\end{tabular} 
& Closed \\ 
\hline
Q10 & If yes, how? If not, why? & - & Open \\ 
\hline
\end{tabular*}
\end{table*}

\pagebreak
\textbf{Section 4} evaluates the framework's completeness and accuracy (see Table \ref{tab:assessment}). It contains nine questions, blending open and closed formats. Participants are asked to rate the framework on scales of completeness and accuracy, suggest potential improvements, and assess the relevance of practices and effects presented in the framework. This section aims to refine the framework based on practitioner feedback. 

\begin{table*}[htb]
\centering
\small
\renewcommand{\arraystretch}{1.3} 
\caption{Questions to assess completeness, accuracy, and usefulness of the framework}
\label{tab:assessment}
\begin{tabular*}{\textwidth}{p{1cm}p{9cm}p{3cm}p{2cm}}
\hline
\textbf{No} & \textbf{Question (Q) Description} & \textbf{Options} & \textbf{Type} \\ 
\hline
\multicolumn{4}{l}{\textbf{Section 4. Assess the framework in relation to its completeness and accuracy}} \\ 
\hline
Q11 & In your opinion, and considering a 0-10 scale, with 0 representing very low completeness and 10 representing high completeness, do you think the framework is complete to represent empathy in software engineering? 
& 0-10 scale & Closed \\ 
\hline
Q12 & What changes does the framework need to better represent empathy in software engineering? 
& - & Open \\ 
\hline
Q13 & In your opinion, and considering a 0-10 scale, with 0 representing very low accuracy and 10 representing high accuracy, do you think the framework is accurate to represent empathy in software engineering? 
& 0-10 scale & Closed \\ 
\hline
Q14 & What changes do you think the framework needs to improve its accuracy? 
& - & Open \\ 
\hline
Q15 & Considering the practices captured in our framework, would they influence any of your software engineering activities? 
& \begin{tabular}[t]{@{}l@{}}
Yes \\ 
No 
\end{tabular} 
& Closed \\ 
\hline
Q16 & If yes, how? If not, why? 
& - & Open \\ 
\hline
Q17 & In your opinion, are the effects of practicing empathy feasible to achieve by the practices captured in our framework? 
& \begin{tabular}[t]{@{}l@{}}
Yes \\ 
No 
\end{tabular} 
& Closed \\ 
\hline
Q18 & If yes, how? If not, why? 
& - & Open \\ 
\hline
Q19 & Do you suggest adding any additional practice to our framework? If yes, please explain the practice and its effects. 
& - & Open \\ 
\hline
\multicolumn{4}{l}{\textbf{Section 5. Assess the structure and use of the framework}} \\ 
\hline
Q20 & Concerning the framework, choose the option that best represents your opinion: 

\begin{itemize}
    \item (a)  I find the framework easy to use to practice empathy in SE.
    \item (b)  I find the framework useful for practicing empathy in SE. 
    \item (c) Assuming the framework was available at my job, I predict I would use it regularly.
    \item (d) Assuming the framework was available at my job, I would recommend it to other practitioners. 
\end{itemize}
& \begin{tabular}[t]{@{}l@{}}
Strongly disagree \\ 
Disagree \\ 
Neutral \\ 
Agree \\ 
Strongly agree 
\end{tabular} 
& Closed \\ 
\hline
Q21 & Considering the usefulness of the framework, how would you use it in your work activities? 
& - & Open \\ 
\hline
\end{tabular*}
\end{table*}

\textbf{Section 5} shifts focus to the framework's structure and applicability in professional contexts (see  Table \ref{tab:assessment}). With two questions, this section uses a mix of closed-ended ratings and open-ended feedback to assess the framework's ease of use, potential adoption in the workplace, and recommendations to peers. It aims to evaluate the practical utility of the framework in real-world SE activities.

 \pagebreak

\textbf{Section 6} examines how the framework can mitigate empathy-related challenges in the workplace. It mixes closed-ended ratings and open-ended feedback that ask participants to identify barriers to empathy they believe can be overcome, describe particularly challenging barriers, and suggest practices from the framework to address them (see Table \ref{tab:survey-barriers}). The goal is to understand how the framework can help mitigate those barriers. 

Finally, \textbf{Section 7} concludes the survey with one open-ended question soliciting additional feedback or comments about the framework (see Table \ref{tab:survey-barriers}). This section ensures participants can share insights not covered in previous questions, providing valuable input for refining the framework.

\begin{table*}[ht]
\centering
\small
\renewcommand{\arraystretch}{1.3} 
\caption{Questions to evaluate challenges and get feedback}
\label{tab:survey-barriers}

\begin{tabular*}{\textwidth}{p{1cm}p{6cm}p{6cm}p{2cm}}
\hline
\textbf{No} & \textbf{Question (Q) Description} & \textbf{Options} & \textbf{Type} \\ 
\hline
\multicolumn{4}{l}{\textbf{Section 6. How the framework can mitigate empathy-related challenges}} \\ 
\hline
Q22 & Which of these barriers do you believe can be overcome in your work environment? 
& \begin{tabular}[t]{@{}l@{}}
Excessive technical focus  \\ 
Individualistic behavior \\ 
Toxic organizational culture \\ 
Challenges in expressing emotions \\ 
Workplace bias \\ 
Work constraints \\ 
None 
\end{tabular} 
& Closed \\ 
\hline
Q23 & How would you approach or address those barriers in your workplace? 
& - & Open \\ 
\hline
Q24 & Which barriers do you consider particularly difficult or impossible to overcome in your work environment? 
& \begin{tabular}[t]{@{}l@{}}
Excessive technical focus  \\ 
Individualistic behavior \\ 
Toxic organizational culture \\ 
Challenges in expressing emotions \\ 
Workplace bias \\ 
Work constraints \\ 
None 
\end{tabular} 
& Closed \\ 
\hline
Q25 & Why do you think these barriers persist despite efforts to address them? 
& - & Open \\ 
\hline
Q26 & Do the practices listed in the framework help overcome those barriers? 
& \begin{tabular}[t]{@{}l@{}}
Yes \\ 
No 
\end{tabular} 
& Closed \\ 
\hline
Q27 & If yes, how? If not, why? 
& - & Open \\ 
\hline
\multicolumn{4}{l}{\textbf{Section 7. Collect additional information and feedback}} \\ 
\hline
Q28 & Do you have any additional comments or feedback regarding the framework that you would like to share? 
& - & Open \\ 
\hline
\end{tabular*}

\end{table*}